\documentclass{IEEEoj}
\usepackage{cite}
\usepackage{amsmath,amssymb,amsfonts,amsthm}

\newtheorem{definition}{Definition}

\usepackage{algorithmic}
\usepackage{graphicx,color}
\usepackage{subcaption}
\usepackage{textcomp}
\usepackage{booktabs}
\usepackage{url}
\usepackage{float}
\usepackage{array}
\usepackage{multirow}
\usepackage{makecell}
\usepackage{tabularx}
\usepackage{verbatim}
\usepackage{cleveref}
\usepackage{cases}
\usepackage{tikz}
\usetikzlibrary{shapes.geometric, arrows.meta, positioning, fit, backgrounds, calc,
                decorations.pathreplacing, spline}
\usepackage{xcolor}

\def\BibTeX{{\rm B\kern-.05em{\sc i\kern-.025em b}\kern-.08em
    T\kern-.1667em\lower.7ex\hbox{E}\kern-.125emX}}

\AtBeginDocument{\definecolor{ojcolor}{cmyk}{0.93,0.59,0.15,0.02}}
\def\OJlogo{\vspace{-4pt}\hskip-4pt\includegraphics[height=18pt]{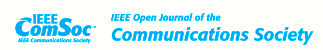}}

\begin{document}

\receiveddate{Date Month, Year}
\reviseddate{Date Month, Year}
\accepteddate{Date Month, Year}
\publisheddate{}
\currentdate{Date Month, Year}
\doiinfo{OJCOMS.2026.XXXXXX}

\title{Performance Analysis of Dynamic Equilibria in Joint Path Selection and Congestion Control in Path-Aware Networks}

\author{Sina Keshvadi\IEEEauthorrefmark{1} \IEEEmembership{}}

\affil{Department of Engineering, Thompson Rivers University, Kamloops, BC V2C 0C8, Canada}

\corresp{Corresponding author: Sina Keshvadi (e-mail: skeshvadi@tru.ca).}

\markboth{Performance Analysis of Dynamic Equilibria in Joint Path Selection}{Keshvadi \textit{et al.}}

\begin{abstract}

  Path-aware networking (PAN) architectures such as SCION, together with emerging high path-diversity systems such as LEO satellite constellations like Starlink, are fundamentally reshaping the Internet by exposing endpoints to tens or hundreds of alternative routes. As multipath transport protocols like MPQUIC approach final standardization, the large-scale deployment of greedy, uncoordinated path selection introduces a critical and poorly quantified threat: self-interested agents migrating toward the instantaneously least-loaded path can trigger persistent, high-amplitude load oscillations that undermine the stability these advanced architectures are designed to provide.

  In this paper, we develop a discrete-time axiomatic framework to analyze the joint dynamics of loss-based congestion control and greedy path selection. We derive the system's long-term dynamic equilibria, which manifest as stable periodic orbits that we term $P$-step oscillations. Our analysis establishes a fundamental and unavoidable design trade-off: higher agent responsiveness improves fairness among competing flows but necessarily degrades network efficiency and convergence. Conversely, we demonstrate that efficiency, convergence, and loss avoidance can be simultaneously maximized at a critical lossless operating point.

  Most significantly, we reveal a sharp phase transition with increasing path diversity. While migration successfully de-synchronizes traffic in high-diversity environments, realistic limited-visibility constraints where agents can probe only a small subset of paths prevent identification of the global minimum. As a result, coherent temporal oscillations are replaced by persistent spatial load imbalance rather than being eliminated. These findings yield design guidelines for building robust multipath transport protocols over the future path-aware Internet.
\end{abstract}

\begin{IEEEkeywords}
  Axiomatic Method,
  Congestion Control,
  Dynamic Equilibria,
  Path Selection,
  Performance Analysis,
  Network Oscillation
\end{IEEEkeywords}

  \maketitle
\section{Introduction}~\label{sec::introduction}
\IEEEPARstart{T}{he} modern Internet is undergoing a profound architectural shift. 
At the network layer, path-aware networking (PAN) architectures such as SCION, together with large-scale LEO satellite constellations such as Starlink, are moving from research prototypes into large-scale deployments~\cite{chuat2022complete,anapaya-deployment}. At the transport layer, multipath extensions to QUIC (MPQUIC) are approaching final standardization \cite{ietf-quic-multipath}. Together, these technologies promise unprecedented path diversity by routinely exposing endpoints to tens or hundreds of verifiable, disjoint routes \cite{herschbach2025path,krahenbuhl2024glids}, along with the ability to exploit that diversity for higher throughput, resilience, and performance.

Yet this convergence introduces a subtle but serious stability risk that remains poorly quantified. When large populations of self-interested agents independently and greedily select paths based on instantaneous load, their uncoordinated migrations can trigger persistent, high-amplitude load oscillations. This phenomenon, long recognized in the selfish routing literature \cite{roughgarden2002bad}, has received far less attention in the concrete setting of loss-based congestion control operating over high-diversity path-aware networks. As MPQUIC deployments scale and SCION-like architectures proliferate, understanding the quantitative performance consequences of this instability becomes essential.

Prior research has largely focused on \emph{preventing} oscillations through centralized traffic engineering, randomization, or inertia mechanisms \cite{khalili2013mptcp,raiciu2011improving}. While these suppression techniques are valuable, they leave a critical gap: we still lack a rigorous understanding of what happens to network efficiency, fairness, convergence, and loss when oscillations \emph{do} occur, particularly as the number of available paths grows into the regime characteristic of real PAN deployments. 
Existing multipath protocols such as MPTCP and MPQUIC, as well as most analytical models, were designed for limited path diversity, typically supporting only a small number of paths (e.g., WiFi combined with LTE, or IPv4 with IPv6). They are not equipped to handle the massive path diversity now offered by path-aware architectures such as SCION and emerging LEO satellite constellations like Starlink. Furthermore, many models rely on fluid approximations that obscure the discrete, stochastic nature of agent migration and window dynamics.

In this paper, we address this gap by developing a discrete-time axiomatic framework for the joint dynamics of loss-based congestion control and greedy path selection. We derive the system's long-term behavior and show that it converges to stable periodic orbits, which we term $P$-step oscillations. Our analysis yields three central contributions.

First, we prove a fundamental and unavoidable design trade-off: increasing agent responsiveness improves fairness among competing flows but necessarily degrades efficiency and convergence speed. Second, we demonstrate that efficiency, convergence, and loss avoidance are not inherently in conflict, and they can be simultaneously optimized at a precisely tuned critical lossless operating point. Third, and most significantly, we uncover a sharp phase transition as path diversity increases. While path migration successfully de-synchronizes traffic in high-diversity environments, realistic constraints on path visibility, where agents can probe only a bounded subset of paths, prevent agents from identifying the global minimum. As a result, coherent temporal oscillations are replaced by persistent spatial load imbalance rather than being eliminated. This finding reveals that massive path diversity does not automatically solve instability. It merely transforms its character.

We validate these theoretical results through discrete-event simulation. The results provide clear and precise guidance for protocol designers on how to balance responsiveness, efficiency, and stability when deploying multipath transport over high-diversity path-aware networks. 

The remainder of this paper is organized as follows.
Section~\ref{sec::background-and-related-work} reviews the necessary background on the axiomatic method and related work in path selection.
Section~\ref{sec::model} formally introduces our discrete-time model for path-aware dynamics and outlines our core assumptions.
In Section~\ref{sec::equilibrium}, we derive the system's dynamic equilibria, identifying the stable oscillation patterns that emerge.
Section~\ref{sec::axioms_intro} defines the five performance axioms used to evaluate these equilibria.
Section~\ref{sec::insights} presents our primary theoretical findings, including the fundamental design trade-offs and the de-synchronization effect.
We validate these theoretical predictions via discrete-event simulation in Section~\ref{sec::evaluation}.
Finally, Section~\ref{sec::discussion} discusses the broader design implications for future protocols, and Section~\ref{sec::conclusion} concludes the paper.

\section{Background and Related Work}\label{sec::background-and-related-work}

\subsection{Background}
The axiomatic method provides a rigorous and principled framework for evaluating systems that must simultaneously satisfy multiple, often competing design objectives. Its philosophical foundations lie in classical economics and social choice theory, where axiomatic frameworks have long been used to formally characterize the properties of desirable (``axioms'') of ``fair'' or ``desirable'' outcomes in systems involving self-interested agents \cite{nash1950bargaining}. Rather than optimizing a single metric or relying solely on empirical evaluation, the axiomatic approach seeks to identify which combinations of properties are fundamentally achievable and which are inherently in conflict.

In the networking domain, the axiomatic approach was first popularized by Lev et al. \cite{lev2016axiomatic}, who demonstrated that abstract properties such as robustness and scale invariance could uniquely characterize classic routing algorithms. Their work proved that any protocol satisfying a specific set of axioms must necessarily be a shortest-path algorithm. Zarchy et al. \cite{zarchy2019axiomatizing} later adapted the methodology to congestion control, defining four foundational axioms (Efficiency, Fairness, Convergence, and Loss Avoidance) and formally proving that certain combinations of these goals are mutually exclusive. Their work established that extreme TCP-friendliness and rapid bandwidth utilization, for example, cannot be achieved simultaneously.

The axiomatic method proceeds in three clearly defined steps~\cite{lev2016axiomatic}. 
First, a mathematical model is constructed that captures the essential interactions and decision rules of the agents under study. 
Second, the model is analyzed to identify its long-term equilibrium state, which represents the stable pattern of behavior to which the system dynamics converge.
Third, the properties of this equilibrium are evaluated against a set of formal axioms that encode desired system characteristics. 
This process reveals the fundamental boundaries of the design space and provides strong impossibility or possibility results that hold independently of specific parameter choices or implementations.

Most recently, Scherrer et al. \cite{scherrer2022axiomatic} extended the axiomatic framework to the joint dynamics of congestion control and path selection. They introduced Responsiveness as a fifth axiom and analyzed the performance implications of end-host-driven path selection. Their work forms the direct foundation of this paper.
In this work, we significantly extend their framework to the high path-diversity regime that characterizes real path-aware networks such as SCION and emerging LEO satellite systems. In particular, we focus on realistic constraints on path visibility and derive new theoretical results regarding dynamic equilibria and performance trade-offs that are directly applicable to the design of MPQUIC and future multipath protocols over path-aware infrastructures.

\subsection{Related Work}

The shift toward vesting path-selection intelligence in network endpoints is reshaping diverse networking domains. In massive-scale data centers, source-driven traffic management is now standard for minimizing flow completion times \cite{alizadeh2014conga}. Emerging LEO satellite constellations rely on source-based mechanisms to adapt to rapid topology changes \cite{handley2018delay}, while proposals for multi-criteria routing allow applications to optimize for complex objectives beyond simple latency \cite{sobrinho2001algebra}. Most notably, in next-generation inter-domain architectures like SCION, end-host path control is foundational, enabling explicit trust and resilience against routing anomalies \cite{chuat2022complete}. This transfer of agency to the edge, however, reintroduces the classic stability challenge: uncoordinated, greedy decisions by self-interested agents can trigger persistent load oscillations.

Existing literature addresses this instability primarily through two oscillation-suppression paradigms. Centralized approaches, such as SDN-based traffic engineering, prevent greedy behavior by computing and installing stable global traffic distributions \cite{jain2013b4}. End-host approaches, conversely, modify agent logic to induce stability without coordination, often by introducing inertia or randomization into path switching \cite{khalili2013mptcp,raiciu2011improving}. While effective at preventing instability, these works do not quantify the performance characteristics of the system when oscillations do occur. Our work takes a fundamentally different approach by analyzing the quantitative impact of oscillations rather than suppressing them.

Recent measurement studies have provided valuable insights into path dynamics in deployed path-aware networks. Herschbach et al. \cite{herschbach2025path} conducted extensive measurements on the global SCIONLab testbed, revealing high path diversity, significant control-plane churn, and short path lifetimes. Krüger et al. \cite{kruger2023path} proposed mechanisms to help applications make more informed path selections in SCION. While these works are important, they focus on empirical characterization and do not analyze the dynamic equilibria that emerge from the interaction between greedy path selection and loss-based congestion control.
Similarly, the rapid growth of LEO satellite constellations such as Starlink has motivated research into multipath routing over highly dynamic topologies \cite{wang2023high}. However, the stability implications of uncoordinated end-host path selection in these high-diversity environments remain largely unexplored.

From a methodological standpoint, prior efforts to model these dynamics typically sacrifice either congestion detail or analytical generality. Foundational work in selfish routing uses the Wardrop model to analyze static traffic distributions and their efficiency (Price of Anarchy) \cite{roughgarden2002bad,correa2004selfish}, but these models cannot capture the window-based dynamics of modern congestion control. Conversely, applied research in Multipath TCP (e.g., LIA and OLIA) captures oscillatory phenomena but relies heavily on heuristic-driven design and experimental validation, limiting its theoretical reach \cite{khalili2013mptcp}.
The axiomatic method has been successfully applied to routing \cite{lev2016axiomatic} and congestion control \cite{zarchy2019axiomatizing}. Scherrer et al. \cite{scherrer2022axiomatic} were the first to extend it to the joint dynamics of path selection and congestion control. Our work builds directly on theirs but significantly extends the analysis to the high path-diversity regime with limited visibility that characterizes both SCION and LEO satellite networks.

In summary, prior work has either focused on preventing oscillations, provided empirical measurements, or developed models that do not jointly capture discrete migration and window dynamics at scale. Our contribution is the first to derive and analyze the dynamic equilibria of the joint system under realistic constraints, revealing both fundamental trade-offs and a previously unknown phase transition in stability as path diversity increases which provides the theoretical foundation needed to design robust multipath protocols for the emerging path-aware Internet.

\section{Modeling Path-Aware Dynamics}~\label{sec::model}

\subsection{System Setup and Assumptions}
We model a path-aware network in which a population of end-host agents compete for bandwidth across multiple disjoint paths. 
Let \(A\) be a set of \(N\) agents, each representing an end-host flow, that compete for bandwidth across a set \(\Pi\) of \(P\) parallel, resource-disjoint bottleneck paths. We assume the total network capacity \(C\) is distributed homogeneously across the paths, so that each path \(\pi \in \Pi\) has identical bottleneck capacity \(C_{\pi} = C/P\).

Each agent \(i \in A\) follows a common Multi-Path Congestion Control (MPCC) protocol, defined by the tuple \((CC, \rho, \sigma)\), where \(CC\) is the underlying congestion control algorithm, \(\rho \in (0,1]\) is the agent's \textit{Responsiveness} (the probability that the agent migrates to a better path in any given time step), and \(\sigma \in [0,1]\) is the \textit{Reset Softness} (the multiplicative factor applied to the agent's congestion window immediately after a path switch). The core notation used throughout this paper is summarized in Table~\ref{tab:notation}.

\begin{table}[!htbp]
  \centering
  \footnotesize
  \caption{Core notation for the analytical model.}
  \label{tab:notation}
  \begin{tabular}{@{}ll@{}}
    \toprule
    \textbf{Symbol} & \textbf{Description} \\
    \midrule
    \(A\)             & Set of \(N\) agents in the network \\
    \(A_\pi(t)\)      & Set of agents utilizing path \(\pi\) at time \(t\) \\
    \(a_\pi(t)\)      & Number of agents on path \(\pi\) at time \(t\), i.e., \(|A_\pi(t)|\) \\
    \(\alpha(\tau)\)  & Additive increase function based on continuity time \(\tau\) \\
    \(C\)             & Total bottleneck capacity of the network \\
    \(C_\pi\)         & Bottleneck capacity of path \(\pi\) \\
    \(\gamma\)        & Multiplicative decrease factor after a loss event \\
    \(L_\pi(t)\)      & Total load on path \(\pi\) at time \(t\) \\
    \(M_\pi(t)\)      & Set of agents migrating away from path \(\pi\) at time \(t\) \\
    \(N\)             & Total number of agents in the network \\
    \(P\)             & Total number of parallel paths in the network \\
    \(\Pi\)           & Set of \(P\) paths in the network \\
    \(\pi_{\min}(t)\) & Path with minimum utilization at time \(t\) \\
    \(\rho\)          & Responsiveness: probability an agent migrates in a time step \\
    \(\sigma\)        & Reset Softness: window scaling factor after migration \\
    \(\tau_i(t)\)     & Continuity time of agent \(i\) at time \(t\) \\
    \(w_i(t)\)        & Congestion window size of agent \(i\) at time \(t\) \\
    \(z(\cdot)\)      & Extrapolation factor for estimating incoming load \\
    \bottomrule
  \end{tabular}
\end{table}

To isolate the fundamental mechanisms driving instability, we make three core assumptions that reflect both worst-case conditions and real-world protocol behavior:
First, agents are myopic and always migrate to the path with the lowest instantaneous utilization. This greedy, load-adaptive behavior, while individually rational, is a well-known driver of herding dynamics and persistent oscillations in competitive routing systems~\cite{roughgarden2002bad}.
Second, when an agent switches paths, it migrates its entire traffic flow at once rather than splitting traffic gradually. This maximizes the impact of each migration event and is consistent with the behavior of many real-world multipath schedulers, which tend to concentrate traffic on the single best-performing path at any given time~\cite{raiciu2011improving}.
Third, all paths are resource-disjoint and have identical bottleneck capacities and round-trip times. This ensures that all agents receive identical performance feedback and therefore react in a highly synchronized manner, creating a worst-case scenario for coherent, network-wide oscillations.

These assumptions are well-aligned with the operational characteristics of next-generation architectures. Production SCION deployments routinely expose tens to hundreds of paths~\cite{herschbach2025path}, and standard multipath schedulers (e.g., in MPQUIC) often approximate our sequential migration model by directing the majority of traffic onto the lowest-latency path.

\subsection{Agent Migration and Load Dynamics}\label{sec::migration}

Having established the basic components of the model, we now describe how agents dynamically select paths and how this selection process affects both the distribution of agents and the load on each path over time.

At every discrete time step \(t\), each agent observes the current utilization of all paths and identifies the path with the minimum instantaneous load, denoted \(\pi_{\min}(t)\). This path is considered the most attractive destination. On every other path \(\pi \neq \pi_{\min}(t)\), each agent \(i \in A_{\pi}(t)\) independently decides whether to migrate to \(\pi_{\min}(t)\) with probability \(\rho\), which we refer to as the agent's responsiveness. Agents that decide to migrate form a random set \(M_{\pi}(t)\) for each suboptimal path.

This migration process has two immediate effects. First, the number of agents on each suboptimal path decreases by the number of agents that chose to leave. Second, the load on those paths is reduced by the sum of the congestion windows carried by the departing agents. Simultaneously, the destination path \(\pi_{\min}(t)\) receives all the migrating agents. Each arriving agent contributes its congestion window scaled by the reset softness factor \(\sigma\), reflecting the fact that the agent resets (or partially resets) its window upon switching paths. In addition, all agents already present on the destination path continue to increase their windows according to the congestion control algorithm.
The complete stochastic evolution of the system is captured by the following recursive equations:

\begin{scriptsize}
  \begin{subequations}
    \label{eq:stochastic}
    \begin{align}
      a_{\pi}(t+1) &=
      \begin{cases}
        a_{\pi}(t) - |M_{\pi}(t)| & \text{if } \pi \neq \pi_{\min}(t) \\
        a_{\pi}(t) + \sum_{\pi' \neq \pi} |M_{\pi'}(t)| & \text{if } \pi = \pi_{\min}(t)
      \end{cases}
      \label{eq:stochastic_agents}
      \\
      L_{\pi}(t+1) &=
      \begin{cases}
        L_{\pi}(t) - \sum_{i \in M_{\pi}(t)} w_i(t) + \sum_{j \in A_{\pi}(t) \setminus M_{\pi}(t)} \Delta w_j(t) & \\\text{if } \pi \neq \pi_{\min}(t) \\
        L_{\pi}(t) + \sum_{\pi' \neq \pi} \sum_{i \in M_{\pi'}(t)} \sigma \cdot w_i(t) + \sum_{j \in A_{\pi}(t)} \Delta w_j(t) & \\ \text{if } \pi = \pi_{\min}(t)
      \end{cases}
      \label{eq:stochastic_load}
    \end{align}
  \end{subequations}
\end{scriptsize}

where \(\Delta w_j(t)\) denotes the change in agent \(j\)'s congestion window as dictated by the underlying congestion control algorithm during the current time step.

While Equations~\eqref{eq:stochastic} provide an exact description of the system, the presence of the random sets \(M_{\pi}(t)\) makes the model analytically intractable and a closed-form equilibrium analysis is therefore not feasible~\cite{scherrer2022axiomatic}. To overcome this difficulty, we adopt a mean-field approximation by replacing all random variables with their expectations. This approximation is well justified when the number of agents \(N\) is large, because the law of large numbers ensures that the aggregate behavior of the system closely follows the evolution of its expected values~\cite{de2003linear}.

Under this approximation, the expected number of agents migrating from any suboptimal path \(\pi\) becomes \(\rho \cdot \hat{a}_{\pi}(t)\), and the expected load carried by these migrants is a proportional share of the path's current load. We further introduce an extrapolation factor \(z(\hat{a}_{\pi}(t)) = (N - \hat{a}_{\pi}(t)) / \hat{a}_{\pi}(t)\) to estimate the total incoming load at the least-utilized path, assuming that load scales proportionally with the number of agents. 
This leads to a fully deterministic system that we analyze in the next subsection. We now turn to the evolution of individual congestion windows.

\subsection{Window Evolution Model}\label{sec::window}

While the migration process governs how agents move between paths, the evolution of each agent's congestion window determines how much traffic it contributes to the path it currently occupies. We now describe a flexible, generic model for window evolution that is expressive enough to capture a wide range of loss-based congestion control algorithms.

At each time step, every agent \(i\) updates its congestion window \(w_i(t)\) according to the following rule. If the total load on its current path does not exceed the path's bottleneck capacity, the agent performs an additive increase. The amount of this increase is governed by a function \(\alpha(\tau_i(t))\), where \(\tau_i(t)\) denotes the agent's continuity time, that is, the number of consecutive time steps the agent has remained on its current path without migrating or experiencing a packet loss. If the path load exceeds capacity, the agent experiences a loss event and multiplicatively decreases its window by a factor \(\gamma \in (0,1)\).
This leads to the following window update equation:

\begin{equation}
\small
\label{eq:window_update}
w_i(t+1) =
\begin{cases}
w_i(t) + \alpha(\tau_i(t)) & \text{if } L_{\pi_i(t)}(t) \leq C_{\pi_i(t)} \quad \text{(no loss)} \\
\gamma \cdot w_i(t) & \text{if } L_{\pi_i(t)}(t) > C_{\pi_i(t)} \quad \text{(loss event)}
\end{cases}
\end{equation}

Consequently, each path contains a heterogeneous population of agents with different continuity times and therefore different window sizes.
The use of continuity time \(\tau\) as an input to the additive increase function allows us to capture the fact that agents which have remained on the same path for a long time tend to have larger windows (and therefore contribute more to path load), while recently migrated agents start with smaller windows after the reset induced by path switching and reset their continuity time to zero.

The formulation in Equation~\eqref{eq:window_update} allows us to represent a broad family of loss-based congestion control protocols by choosing different functional forms for \(\alpha(\tau)\). For example, standard TCP Reno corresponds to the simple case \(\alpha(\tau) = 1\) (constant additive increase per round-trip time). Protocols with slow-start behavior can be modeled by making \(\alpha(\tau)\) grow exponentially with \(\tau\) during the initial phase. More advanced protocols such as CUBIC can be approximated by choosing a non-linear function \(\alpha(\tau)\) that reflects their cubic window growth curve.
This flexibility allows us to derive results that are not tied to one specific congestion control algorithm but instead apply broadly to the entire class of loss-based protocols. It also enables us to study how different increase policies interact with path migration to produce different equilibrium behaviors.
We are now ready to combine the migration dynamics and window evolution into a single deterministic model.

\subsection{Deterministic Mean-Field Model}\label{sec::deterministic}

We now combine the agent migration process and the window evolution model into a single deterministic description of the system. 
The transition to a deterministic model is achieved through a mean-field approximation. Because the number of agents \(N\) is assumed to be large, the law of large numbers implies that the aggregate behavior of the system closely follows the evolution of its expected values~\cite{de2003linear}. In other words, the random fluctuations caused by individual migration decisions average out, and the system can be accurately described by tracking only the expected number of agents and expected load on each path. This mean-field approximation is well-founded for large populations and has been successfully applied in previous work on path selection dynamics~\cite{scherrer2022axiomatic,de2003linear}.

Under the mean-field approximation, the expected number of agents on each path evolves according to the following rule. On any suboptimal path, a fraction \(\rho\) of the agents are expected to migrate away at every time step. On the minimum-utilization path, the expected inflow of agents equals the total expected outflow from all other paths. This leads to the following deterministic recursion for the expected agent count \(\hat{a}_{\pi}(t)\):

\begin{footnotesize}
  \begin{equation}
    \label{eq:expected_agents}
    \hat{a}_{\pi}(t+1) =
    \begin{cases}
      (1 - \rho) \cdot \hat{a}_{\pi}(t) & \text{if } \pi \neq \pi_{\min}(t) \\
      \hat{a}_{\pi}(t) + \rho \bigl(N - \hat{a}_{\pi}(t)\bigr) & \text{if } \pi = \pi_{\min}(t)
    \end{cases}
  \end{equation}
\end{footnotesize}

On suboptimal paths, the expected load decreases due to departing agents but increases due to the additive increase of the agents that remain. On the destination path, the expected load increases due to arriving agents (whose windows are scaled by \(\sigma\)) as well as the additive increase of all agents already present. Additionally, if the load on a path exceeds its capacity, a loss event occurs and all windows on that path are multiplicatively decreased by \(\gamma\).

\begin{figure}[htbp]
  \centering
  \includegraphics[width=1\columnwidth]{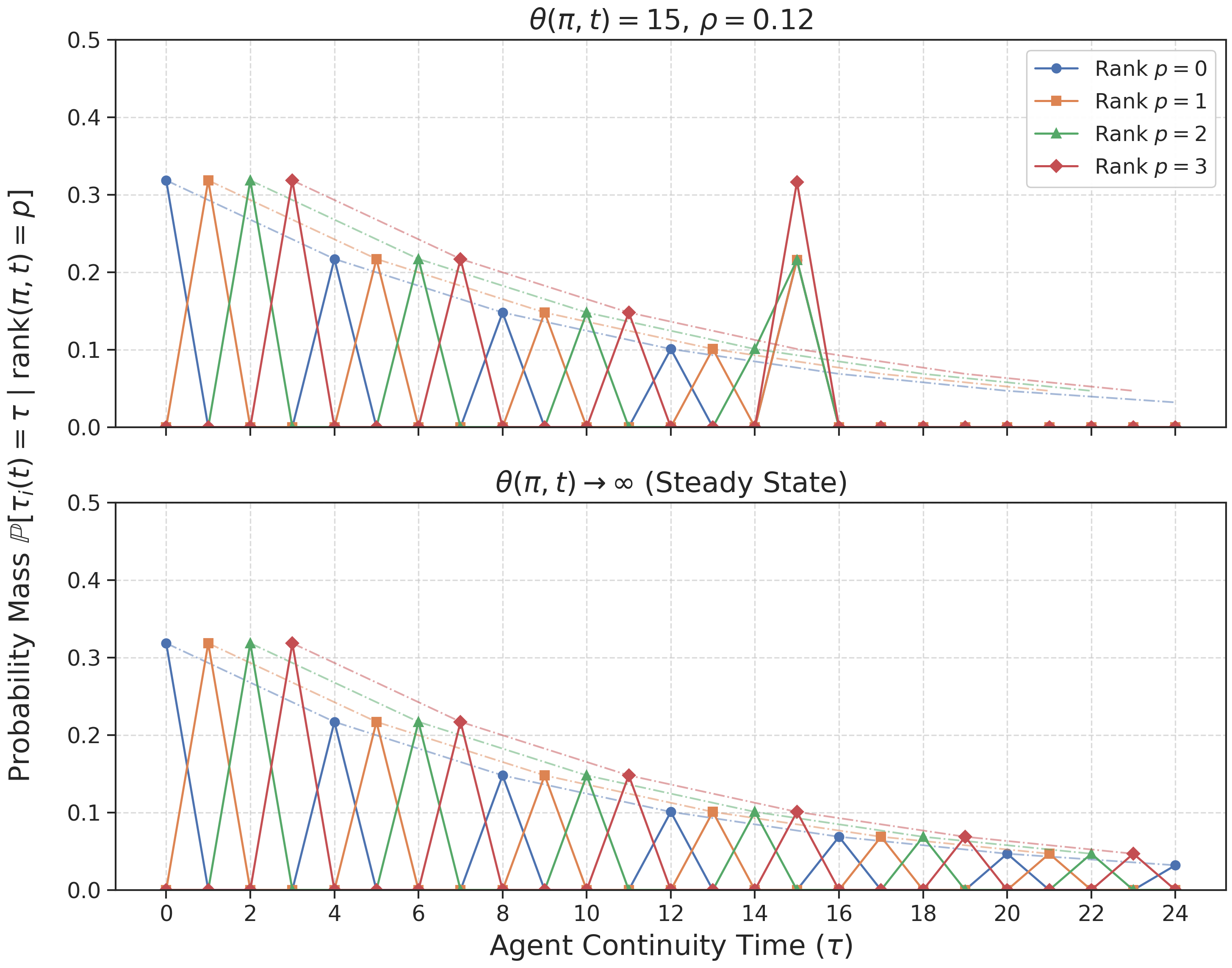}
  \caption{The probability distribution of agent continuity times for a system with \(P=4\) paths, conditioned on a path's rank (\(p=0, 1, 2, 3\)). The periodic influx of new agents to the Rank 0 path (blue) creates a heterogeneous distribution with significant mass at \(\tau=0\). As the path shifts to higher ranks (orange, green, red) in subsequent steps, this mass shifts to higher \(\tau\) values.}
  \label{fig:continuity_dist}
\end{figure}

After incorporating these effects and introducing the extrapolation factor \(z(\hat{a}_{\pi}(t)) = (N - \hat{a}_{\pi}(t)) / \hat{a}_{\pi}(t)\) to estimate incoming load, the deterministic recursion for expected load \(\hat{L}_{\pi}(t)\) takes the following form:

\begin{scriptsize}
\begin{equation}
\label{eq:expected_load}
\hat{L}_{\pi}(t+1) =
\begin{cases}
(1 - \rho) \hat{L}_{\pi}(t) + \hat{\alpha}_{\pi}(t) (1 - \rho) \hat{a}_{\pi}(t) & \text{if } \pi \neq \pi_{\min}(t),\ \hat{L}_{\pi}(t) \le C_{\pi} \\
\bigl(1 + \rho \sigma z(\cdot)\bigr) \hat{L}_{\pi}(t) + \hat{\alpha}_{\pi}(t) \hat{a}_{\pi}(t) & \text{if } \pi = \pi_{\min}(t),\ \hat{L}_{\pi}(t) \le C_{\pi} \\
\gamma (1 - \rho) \hat{L}_{\pi}(t) & \text{if } \pi \neq \pi_{\min}(t),\ \hat{L}_{\pi}(t) > C_{\pi} \\
\bigl(\gamma + \rho \sigma z(\cdot)\bigr) \hat{L}_{\pi}(t) & \text{if } \pi = \pi_{\min}(t),\ \hat{L}_{\pi}(t) > C_{\pi}
\end{cases}
\end{equation}
\end{scriptsize}

where \(\hat{\alpha}_{\pi}(t)\) is the average additive increase per agent on path \(\pi\), computed as the expectation of \(\alpha(\tau)\) over the distribution of continuity times of agents on that path.
This distribution is path-rank dependent: the periodic influx of newly migrated agents onto the minimum-utilization path creates a heterogeneous mixture of continuity times that evolves deterministically as the path cycles through ranks.
Figure~\ref{fig:continuity_dist} shows the probability distribution of continuity times for a representative system with \(P=4\) paths.

Equations~\eqref{eq:expected_agents} and~\eqref{eq:expected_load} together constitute the final deterministic model that forms the basis of all subsequent analysis in this paper. This model captures the essential interplay between greedy path selection, loss-based congestion control, and the resulting dynamic equilibria, while remaining analytically tractable. In the following sections, we will derive the long-term behavior of this system and evaluate its performance properties using the axiomatic framework.

\subsection{Model Scope and Limitations}\label{sec::limitations}

In this subsection, we explicitly discuss the scope of our model, the assumptions we made, and the limitations that necessarily follow from those choices. 

Our model rests on three core assumptions that were chosen to isolate the fundamental mechanisms driving instability while remaining grounded in the operational reality of path-aware networks. The assumption of greedy, load-adaptive selection reflects the behavior of many practical multipath schedulers that attempt to maximize performance for their own flows. While this myopic behavior can lead to herding and oscillations, it is a reasonable approximation of how protocols like MPQUIC currently operate in the absence of sophisticated coordination mechanisms. The assumption of sequential multi-path usage in which agents migrate their entire traffic at once is also well-aligned with real-world scheduler implementations, which often concentrate the vast majority of traffic on the single best-performing path at any given time. Finally, the assumption of homogeneous and disjoint paths creates a worst-case scenario for coherent, network-wide oscillations. While real networks certainly exhibit heterogeneity in capacity and latency, this assumption allows us to derive clean analytical results that reveal the fundamental tension between responsiveness and stability. In practice, path heterogeneity would likely weaken synchronization and reduce oscillation amplitude, making our results a conservative bound on instability.

Nevertheless, several important aspects of real networks are not captured by our model. First, we assume that all paths are resource-disjoint with no shared bottlenecks. In reality, paths often share links deeper in the network, which can introduce correlations in loss events and complicate the migration dynamics. Second, our model assumes synchronous feedback, meaning all agents observe and react to path conditions at the same discrete time steps. Real networks exhibit asynchrony due to different round-trip times and propagation delays, which can desynchronize agent behavior and potentially dampen oscillations. Third, our framework is currently limited to loss-based congestion control. Modern protocols such as BBR~\cite{cardwell2017bbr} and Copa~\cite{arun2018copa} rely on latency and delivery rate signals rather than packet loss. These protocols exhibit fundamentally different dynamics, and extending our axiomatic analysis to model-based and latency-based algorithms represents an important direction for future research.
%

\section{Equilibrium Characterization}\label{sec::equilibrium}

Having derived the deterministic model that governs the expected evolution of agent counts and loads, we now analyze its long-term behavior. A central finding of our analysis is that the system does not converge to a static fixed point. Instead, it enters a stable periodic orbit, a dynamic equilibrium in which the state of the system (the vector of per-path agent counts and loads) evolves through a finite, repeating sequence of values~\cite{strogatz2024nonlinear}.
This behavior arises from the interplay of two coupled forces inherent to greedy migration. The first is the \emph{In-migration Jump}. At every time step, the least-utilized path \(\pi_{\min}(t)\) receives a simultaneous influx of migrating agents from all \(P-1\) other paths. For any non-trivial responsiveness \(\rho > 0\), this sudden additive load is consistently over-reactive. The destination path almost invariably becomes the most heavily loaded path (Rank 0) in the subsequent time step. The second force is the \emph{Out-migration Shift}. Concurrently, every other path experiences an orderly outflow of a fraction \(\rho\) of its agents. Because this reduction is proportional to the current population, it is order-preserving: a path at rank \(p\) will always retain more remaining load than a path at rank \(p+1\).

As a direct consequence of these two forces, the system is forced into a predictable rank permutation: the path at rank \(p \in [0, P-2]\) simply shifts to rank \(p+1\), while the path at rank \(P-1\) shifts to rank 0. This emergent cyclic behavior is known as a \textit{\(P\)-step oscillation}~\cite{scherrer2022axiomatic}.

\begin{definition}[P-Step Oscillation]
A system's expected dynamics exhibit a \(P\)-step oscillation if there exists a time \(t_0\) after which the rank of any path \(\pi \in \Pi\) (ordered from most to least utilized) evolves according to the following rule for all \(T \ge 0\):
\[
\text{If } \operatorname{rank}(\pi, t_0) = p, \quad \text{then } \operatorname{rank}(\pi, t_0 + T) = (p + T) \pmod{P}.
\]
\end{definition}

\begin{figure}[htbp]
	\centering
	\includegraphics[width=1\columnwidth]{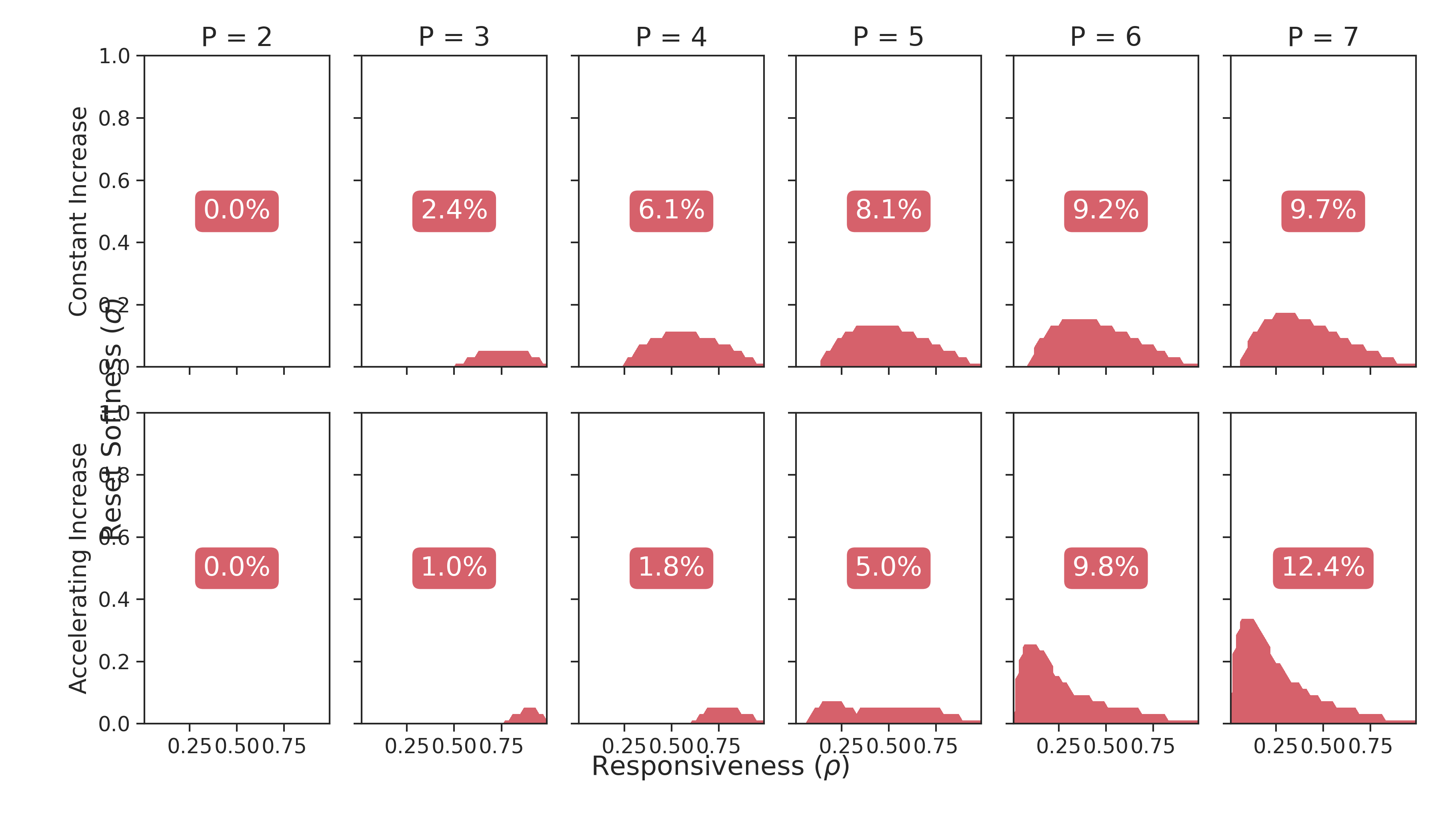}
	\caption{Logical consistency of the $P$-step oscillation across the parameter space. The small shaded regions indicate parameter combinations where the oscillation pattern is inconsistent. The vast unshaded area confirms that the $P$-step oscillation is a robust, fundamental emergent property of the system.}
	\label{fig:pstep_consistency}
\end{figure}

To confirm that the \(P\)-step oscillation is a fundamental property of the system rather than a fragile artifact of specific parameter choices, we verified its logical consistency across the entire parameter space. As illustrated in Figure~\ref{fig:pstep_consistency}, the pattern holds for the vast majority of the parameter space. The only exceptions occur in small, non-physical corner cases (e.g., as \(\rho \to 0\), where migration effectively ceases, or in degenerate cases with extremely small \(N\)). This robustness confirms that the \(P\)-step oscillation is an intrinsic consequence of the model dynamics.

\subsection{Equilibrium Agent Distribution}\label{sec::agent_eq}

We now use the deterministic structure of the \(P\)-step oscillation to derive the equilibrium distribution of agents across ranks. Consider an arbitrary path \(\pi\) that is at rank 0 (most loaded) at some time \(t_0\). Over the next \(P-1\) time steps, this path will only experience agent outflow according to Equation~\eqref{eq:expected_agents} (case 1). In the final step of the cycle, it will receive the full inflow of agents from all other paths (case 2), returning it to rank 0. This cyclic process can be expressed as a first-order difference equation relating the agent count at the beginning of one cycle to the agent count at the beginning of the next cycle:
\begin{footnotesize}
  \begin{equation}
    \label{eq:cycle}
    \hat{a}_{\pi}(t_0 + P) = (1 - \rho)^P \hat{a}_{\pi}(t_0) + \rho N
  \end{equation}
\end{footnotesize}

At equilibrium, the agent count must be periodic with period \(P\). Therefore, the fixed point of Equation~\eqref{eq:cycle} must satisfy \(\hat{a}_{\pi}(t_0 + P) = \hat{a}_{\pi}(t_0)\). Solving this fixed-point equation yields the unique equilibrium agent count for a path at rank \(p\):

\begin{footnotesize}
\begin{equation}
\label{eq:agent_eq}
\hat{a}^{(p)} = \frac{(1 - \rho)^p \cdot \rho \cdot N}{1 - (1 - \rho)^p}
\end{equation}
\end{footnotesize}

This closed-form expression has several important properties. First, it is strictly decreasing in rank \(p\): paths at lower ranks (more loaded) have fewer agents at equilibrium than paths at higher ranks (less loaded). This is intuitive because heavily loaded paths lose agents more rapidly. Second, the distribution becomes more uniform as \(\rho\) increases. When \(\rho\) is close to 1, agents migrate very aggressively, which tends to equalize the number of agents across paths. When \(\rho\) is small, agents exhibit low mobility and the distribution becomes highly skewed, with most agents concentrated on a few lightly loaded paths.

We can also show that the system converges exponentially fast to this equilibrium from any initial condition. The trajectory of the expected agent count on a path starting at rank \(p\) at time \(t_p\) is given by

\[
\hat{a}_{\pi}^{(p)}(t) = \bigl(\hat{a}_{\pi}(t_p) - \hat{a}^{(p)}\bigr) \cdot (1 - \rho)^{t - t_p} + \hat{a}^{(p)}.
\]

As \(t \to \infty\), the first term vanishes, and the system converges to the unique equilibrium distribution \(\hat{a}^{(p)}\) regardless of the initial state.

Having characterized the equilibrium number of agents on each rank, we now turn to the equilibrium load distribution, which determines the amplitude of the resulting oscillations.

\subsection{Equilibrium Load and Oscillation Amplitude}\label{sec::load_eq}

While the equilibrium agent distribution derived in the previous subsection tells us how many agents occupy each rank, it is the load on each path (not the number of agents) that ultimately determines network performance. In this subsection, we derive the expected load on the most loaded path (\(\hat{L}^{(0)}\)) and the least loaded path (\(\hat{L}^{(P-1)}\)). These two quantities fully characterize the amplitude of the \(P\)-step oscillation.

To derive the equilibrium load, we again consider one full cycle of \(P\) steps. Starting from a path at rank \(P-1\) (least loaded), we track how its load evolves as it moves through ranks \(P-2, \dots, 0\) and then returns to rank \(P-1\). During this cycle, the path experiences a combination of agent outflow, window growth, and (in the final step) a large inflow of agents whose windows have been scaled by \(\sigma\).

After substituting the equilibrium agent counts \(\hat{a}^{(p)}\) into the deterministic load recursion and solving the resulting fixed-point equation, we obtain the following closed-form expressions for the equilibrium load on the maximum and minimum rank paths:
\begin{footnotesize}
  \begin{equation}
    \label{eq:load_max}
    \hat{L}^{(0)} = \frac{\bigl((1 + \rho \sigma z(\cdot)) \cdot (\sum_{p=0}^{P-2} \hat{a}^{(p)}) + \hat{a}^{(P-1)}\bigr) \cdot \hat{a}^{(P-1)}}{1 - (1 + \rho \sigma z(\cdot)) \cdot (1 - \rho)^{P-1}}
  \end{equation}
\end{footnotesize}
  
\begin{footnotesize}
\begin{equation}
\label{eq:load_min}
\hat{L}^{(P-1)} = \frac{\bigl(\sum_{p=0}^{P-2} \hat{a}^{(p)} + \hat{a}^{(P-1)}\bigr) \cdot (1 - \rho)^{P-1} \cdot \hat{a}^{(P-1)}}{1 - (1 + \rho \sigma z(\cdot)) \cdot (1 - \rho)^{P-1}}
\end{equation}
\end{footnotesize}

The oscillation amplitude is defined as the difference between these two loads:

\begin{footnotesize}
\[
A = \hat{L}^{(0)} - \hat{L}^{(P-1)}.
\]
\end{footnotesize}

This amplitude is a direct measure of the severity of the instability. A large amplitude indicates that paths experience severe periodic overload and under-utilization, which degrades both efficiency and predictability. Importantly, the amplitude depends on all three protocol parameters: responsiveness \(\rho\), reset softness \(\sigma\), and the number of paths \(P\). As we will show in later sections, there exists a fundamental trade-off: increasing \(\rho\) (more aggressive migration) tends to increase amplitude, while tuning \(\sigma\) (softer resets) can help mitigate it.

The expressions in Equations~\eqref{eq:load_max} and~\eqref{eq:load_min} also allow us to determine whether the equilibrium is \textit{lossless} or \textit{lossy}. If \(\hat{L}^{(0)} \le C_{\pi}\), the system operates in the lossless regime: no path ever exceeds capacity, and no multiplicative decreases occur. If \(\hat{L}^{(0)} > C_{\pi}\), the system enters the lossy regime, where periodic overload triggers window reductions. These two regimes exhibit qualitatively different structural patterns, which we analyze in the next subsection.

\subsection{Lossless vs Lossy Equilibria}\label{sec::lossless_lossy}

The system is in the lossless regime if the peak load on the most loaded path satisfies \(\hat{L}^{(0)} \le C_{\pi}\). In this regime, no path ever exceeds its capacity, and therefore no multiplicative window decreases ever occur. All agents continuously increase their windows according to the additive increase function \(\alpha(\tau)\). The oscillation is purely the result of migration dynamics, and the amplitude is determined entirely by the interplay between \(\rho\), \(\sigma\), and path diversity \(P\).

The system enters the lossy regime when \(\hat{L}^{(0)} > C_{\pi}\). In this regime, the most loaded path periodically exceeds capacity, triggering multiplicative decreases (\(\gamma < 1\)) on all agents present on that path. These loss events introduce an additional source of variability and fundamentally alter the structure of the oscillation. Our analysis reveals that lossy equilibria can exhibit two qualitatively different structural patterns, depending on the value of responsiveness \(\rho\). 

When agents migrate aggressively (high responsiveness, \(\rho \gtrsim 0.3\)), the loss event is absorbed into the regular \(P\)-step cycle. The multiplicative decrease occurs at a predictable point in the cycle, typically when the path is at rank 0, and the rank ordering is preserved throughout. The oscillation remains a clean \(P\)-step pattern, but with reduced amplitude due to the window resets.

\begin{figure}[htbp]
	\centering
	\begin{subfigure}[b]{0.45\textwidth}
		\includegraphics[width=\textwidth]{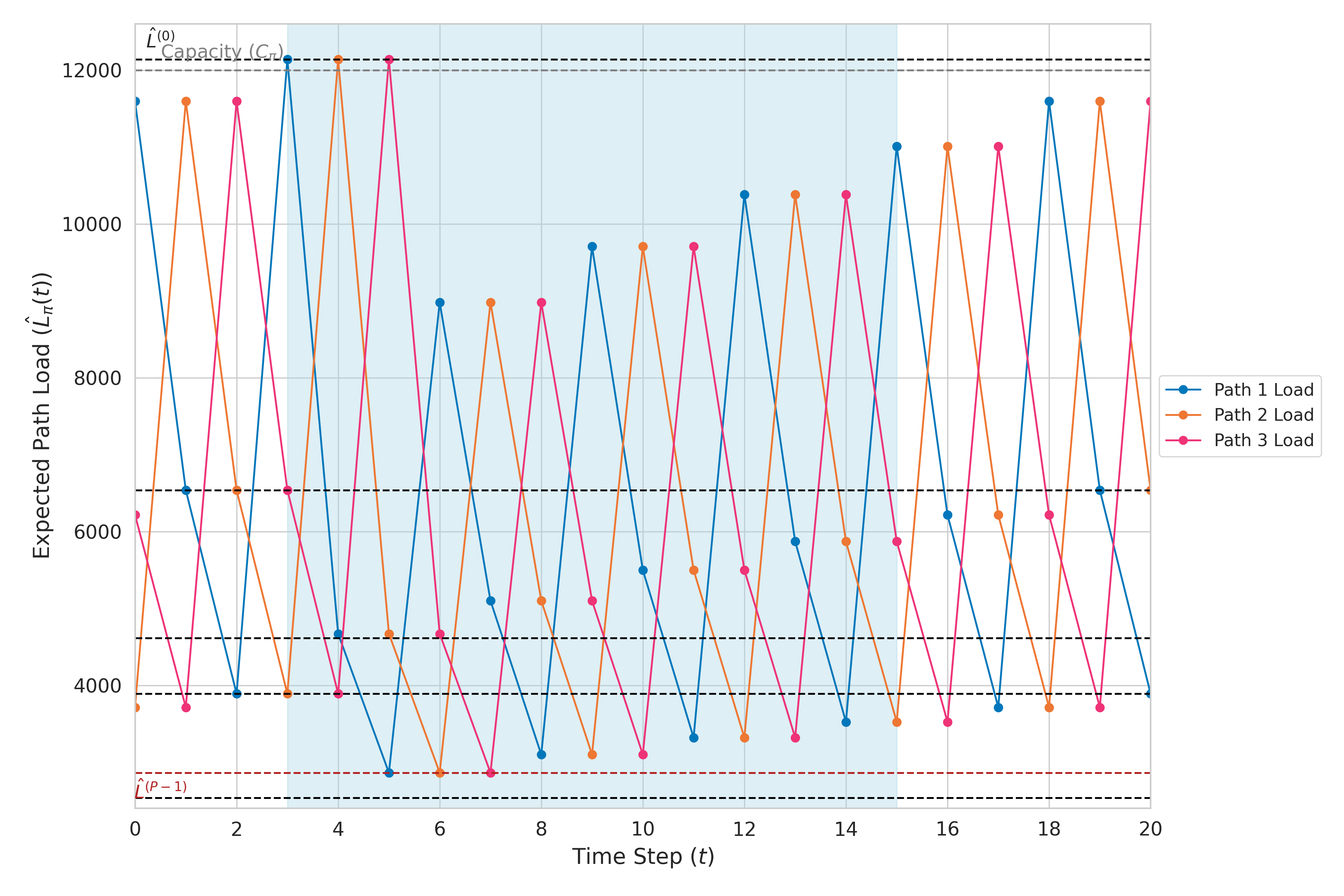}
		\caption{High Responsiveness ($\rho = 0.45$)}
		\label{fig:lossy_high_rho}
	\end{subfigure}
	\hfill
	\begin{subfigure}[b]{0.45\textwidth}
		\includegraphics[width=\textwidth]{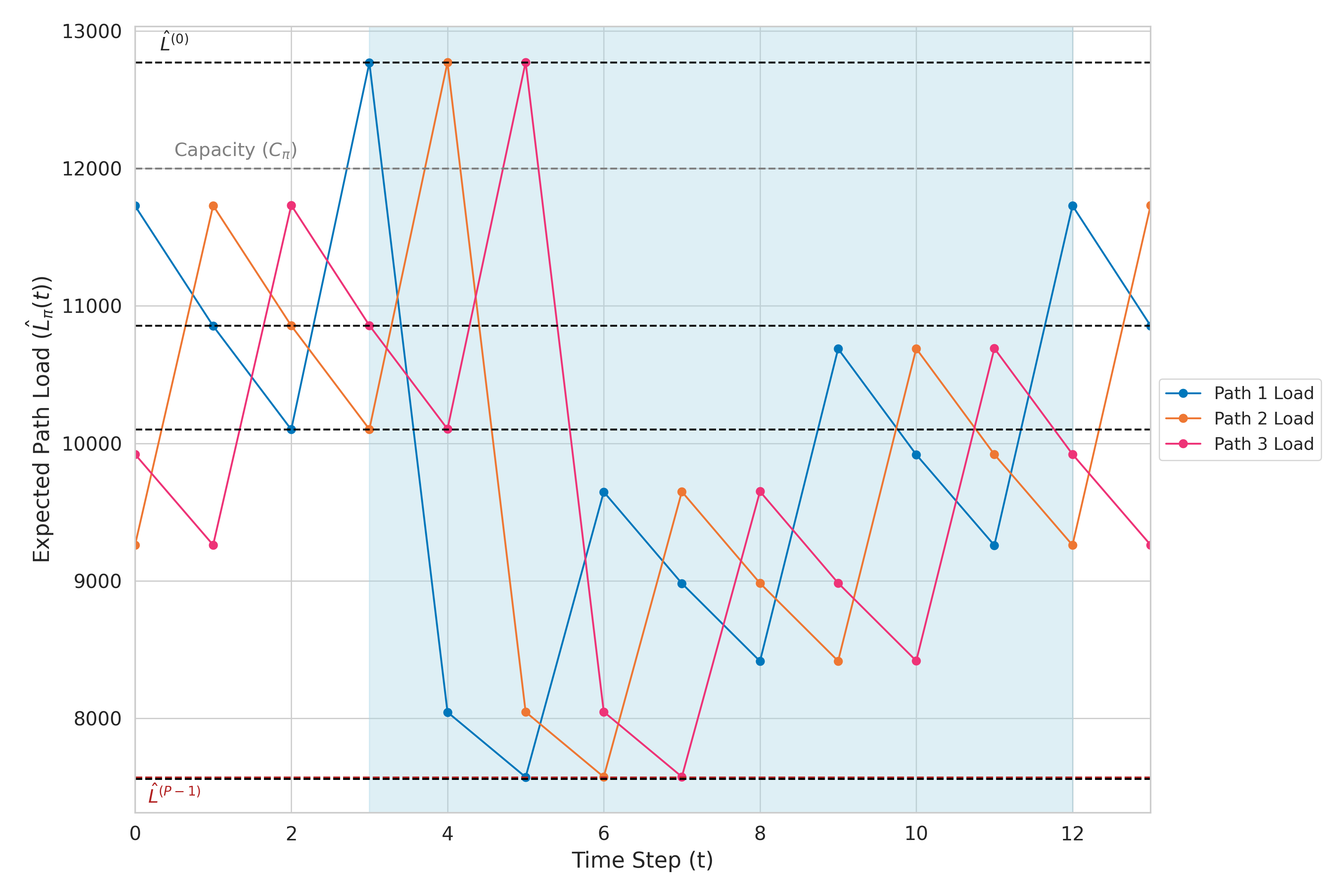}
		\caption{Low Responsiveness ($\rho = 0.1$)}
		\label{fig:lossy_low_rho}
	\end{subfigure}
  \caption{Structural patterns of lossy dynamic equilibria. (a) High Responsiveness ($\rho = 0.45$): loss is absorbed into the regular $P$-step cycle. (b) Low Responsiveness ($\rho = 0.1$): multiplicative decrease temporarily disrupts rank ordering, but the $P$-step cycle immediately resumes after one step.}
	\label{fig:lossy_patterns}
\end{figure}

In contrast, when agents migrate more conservatively (low responsiveness, \(\rho \lesssim 0.2\)), the multiplicative decrease is severe enough to temporarily disrupt the rank ordering. The overloaded path may drop from rank 0 directly to rank \(P-1\) in a single step, breaking the normal cyclic permutation for one time step. After this disruption, the \(P\)-step pattern immediately resumes. This creates a more irregular oscillation with occasional resets in the rank sequence.
These two patterns are illustrated in Figure~\ref{fig:lossy_patterns}.

These two patterns have different implications for performance. The high-responsiveness pattern is more predictable but can still cause significant throughput loss due to frequent window reductions. The low-responsiveness pattern is less predictable and can lead to larger transient imbalances, but the overall frequency of loss events may be lower. Understanding which pattern a protocol operates in is essential for tuning parameters such as \(\rho\) and \(\sigma\) to achieve desired performance.

\subsection{Convergence to Equilibrium}\label{sec::convergence}

We now prove that the system converges exponentially fast to the periodic orbit characterized in the previous subsections. This result is important because it establishes that the equilibria we have derived are not only stable but also practically reachable from any initial state.

Consider a path that starts at rank \(p\) at time \(t_p\). The trajectory of its expected agent count is given by the closed-form expression

\begin{footnotesize}
  \begin{equation}
    \label{eq:trajectory}
    \hat{a}_{\pi}^{(p)}(t) = \bigl(\hat{a}_{\pi}(t_p) - \hat{a}^{(p)}\bigr) \cdot (1 - \rho)^{t - t_p} + \hat{a}^{(p)}.
  \end{equation}
\end{footnotesize}
  
The first term represents the transient deviation from equilibrium, while the second term is the equilibrium value derived in Equation~\eqref{eq:agent_eq}. Because \(0 < \rho \le 1\), the base of the exponential \((1 - \rho)\) is strictly less than 1. Therefore, as \(t \to \infty\), the transient term vanishes exponentially fast with rate \((1 - \rho)\), regardless of the initial deviation \(\hat{a}_{\pi}(t_p) - \hat{a}^{(p)}\).
A similar convergence result holds for the expected load on each path. Although the load recursion is more complex, it is still a linear system with a contraction mapping in the transient component. Consequently, both agent counts and loads converge exponentially to their unique periodic orbits.

\begin{figure}[htbp]
	\centering
	\begin{subfigure}[b]{\columnwidth}
		\includegraphics[width=\textwidth]{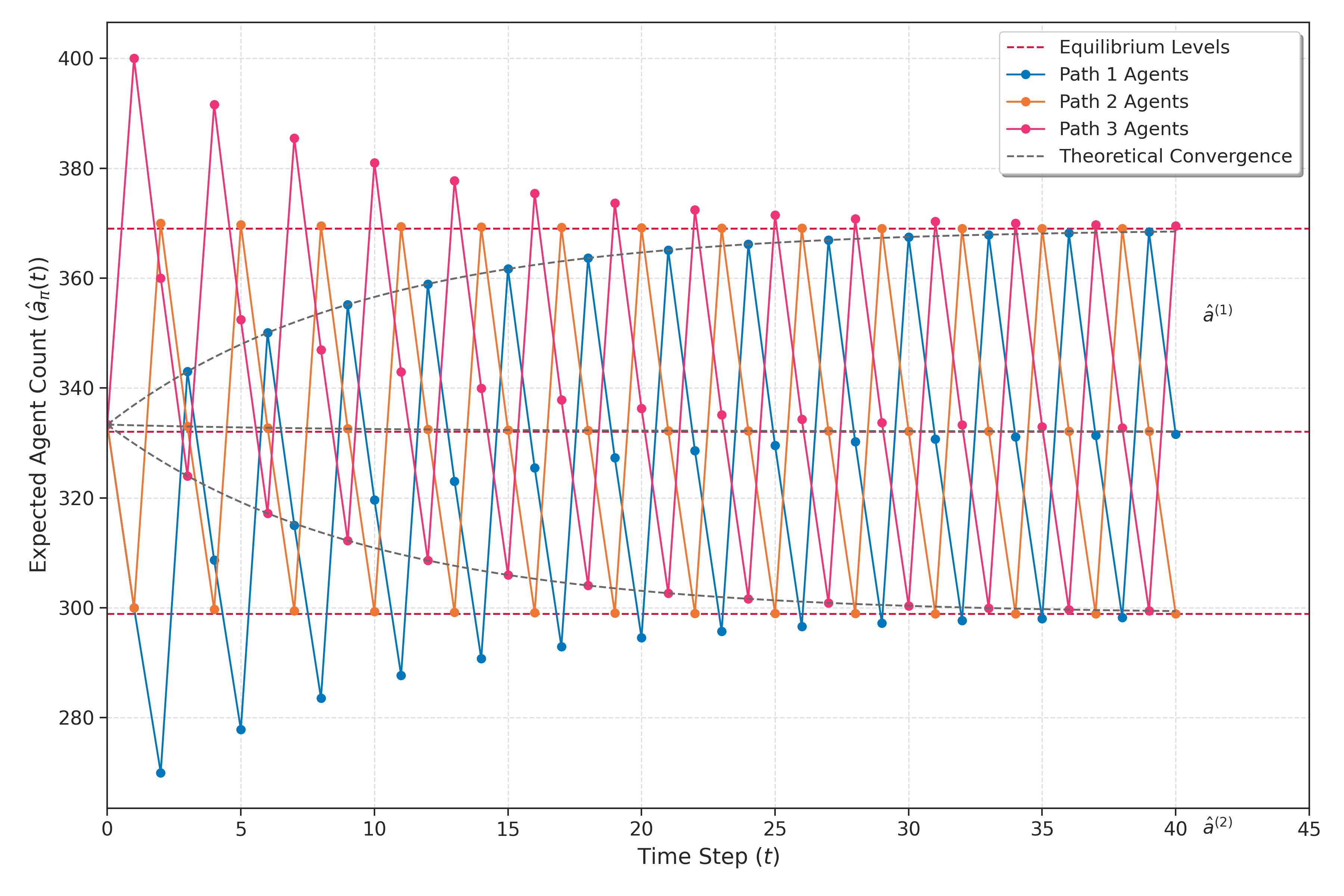}
		\caption{Agent count convergence to the equilibrium distribution.}
		\label{fig:lossless_agent_conv}
	\end{subfigure}
	\hfill
	\begin{subfigure}[b]{\columnwidth}
		\includegraphics[width=\textwidth]{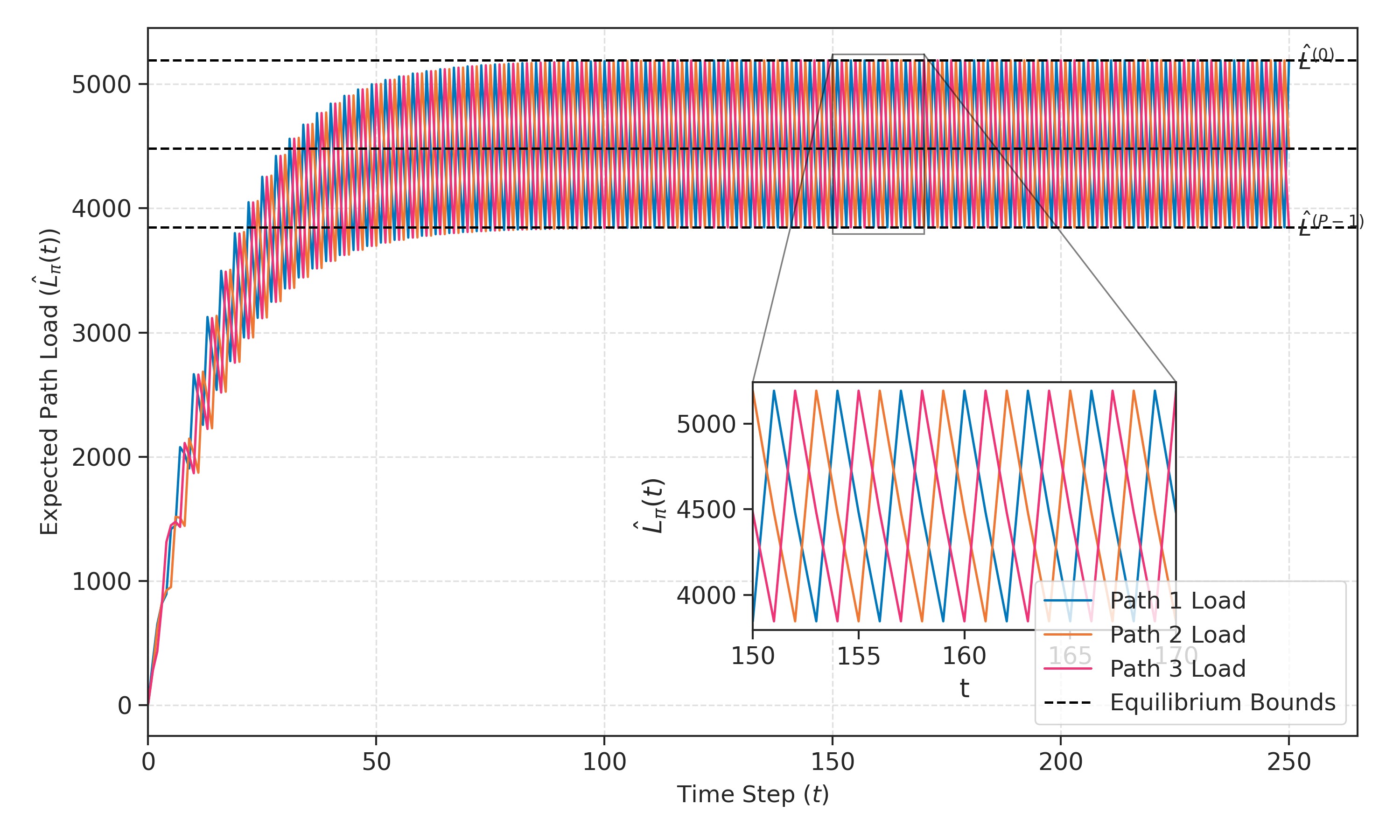}
		\caption{Path load convergence.}
		\label{fig:lossless_flow_conv}
	\end{subfigure}
	\caption{Exponential convergence of the system to the lossless dynamic equilibrium. Both agent counts (top) and path loads (bottom) converge exponentially fast to the unique stable periodic orbit predicted by our analysis, regardless of initial conditions.}
	\label{fig:lossless_convergence}
\end{figure}

As shown in Figure~\ref{fig:lossless_convergence}, numerical simulations confirm this exponential convergence behavior. Both the number of agents on each path and the aggregate load converge rapidly to the stable periodic orbit predicted by our analysis.

This fast convergence has important practical implications. First, it means that the system quickly forgets its initial conditions. Even if the network starts in a highly imbalanced state (e.g., after a sudden surge of new flows or a major path failure), it will rapidly approach the stable oscillation pattern predicted by our analysis. Second, it implies that the equilibria are robust to perturbations. Small disturbances, such as temporary changes in traffic demand or short-term path outages, will not prevent the system from returning to its long-term periodic behavior.

\section{Axiomatic Performance Metrics for Path-Aware Protocols}\label{sec::axioms_intro}

We use the axiomatic approach to evaluate the performance of the dynamic equilibria derived in Section~\ref{sec::equilibrium}. This method defines a set of formal properties (axioms) that a desirable system should satisfy~\cite{lev2016axiomatic,zarchy2019axiomatizing,scherrer2022axiomatic}. By determining which combinations of axioms are simultaneously achievable, we obtain fundamental design trade-offs that hold independently of specific parameter choices.

\subsection{Performance Axioms}\label{sec::axioms_def}

We define five performance axioms that together capture the key dimensions of interest for multipath transport over path-aware networks.

\textit{Efficiency} measures the productive use of network capacity. In a system with periodic oscillations, efficiency is defined as the minimum sustained utilization over one full cycle:

\begin{equation}
\label{eq:efficiency}
\epsilon = \min_{t \ge t_0,\ \pi \in \Pi} \frac{\hat{L}_{\pi}(t)}{C_{\pi}}.
\end{equation}

Higher values of \(\epsilon \in [0,1]\) are desirable.

\textit{Loss Avoidance} quantifies the ability to respect link capacity. We define the loss avoidance rating as the maximum fractional overshoot above capacity during the periodic orbit:

\begin{equation}
\label{eq:loss_avoidance}
\lambda = \max_{t \ge t_0,\ \pi \in \Pi} \left\{ 0,\ \frac{\hat{L}_{\pi}(t) - C_{\pi}}{C_{\pi}} \right\}.
\end{equation}

Lower values of \(\lambda \ge 0\) are desirable, with \(\lambda = 0\) indicating a perfectly lossless protocol.

\textit{Convergence} measures stability and predictability. We define the convergence rating as the ratio of minimum to maximum load over the cycle:

\begin{equation}
\label{eq:convergence}
\gamma_{\rm conv} = \frac{\min_{t \ge t_0,\ \pi \in \Pi} \hat{L}_{\pi}(t)}{\max_{t \ge t_0,\ \pi \in \Pi} \hat{L}_{\pi}(t)}.
\end{equation}

Values of \(\gamma_{\rm conv} \in [0,1]\) closer to 1 indicate lower-amplitude oscillations and higher predictability.

\textit{Fairness} concerns equitable bandwidth allocation. Because agents have heterogeneous histories due to migration and loss events, traditional metrics such as Jain's index fail to capture persistent inequity. We therefore adopt the population-wide variance of congestion window sizes, which directly quantifies the long-term disparity between long-term residents (who accumulate large windows) and frequent migrants (who repeatedly reset to small windows)~\cite{zarchy2019axiomatizing}. We define the fairness rating as

\begin{equation}
\label{eq:fairness}
\eta = \text{Var}_{\pi \in \Lambda} [w_i(t)]_{t \ge t_0},
\end{equation}

where the variance is taken with respect to the stationary distribution of the Markov process governing a single agent's window evolution. Lower values of \(\eta \ge 0\) are desirable, with \(\eta = 0\) indicating perfect equity.

\textit{Responsiveness} captures the agility with which agents exploit better paths. We define the responsiveness rating simply as the per-step migration probability:

\begin{equation}
\label{eq:responsiveness}
\text{Responsiveness} = \rho.
\end{equation}

Higher values of \(\rho \in (0,1]\) correspond to more agile protocols.

\subsection{Evaluating Axioms on Dynamic Equilibria}\label{sec::axioms_eval}

All five axioms are evaluated on the periodic orbits derived in Section~\ref{sec::equilibrium}. Because the system exhibits a stable \(P\)-step oscillation, each axiom is computed over one complete cycle of length \(P\).

For Efficiency, Loss Avoidance, and Convergence, the evaluation follows directly from the closed-form load expressions in Equations~\eqref{eq:load_max} and~\eqref{eq:load_min}. The efficiency rating \(\epsilon\) is the minimum utilization over the cycle, the loss avoidance rating \(\lambda\) is the maximum fractional overshoot, and the convergence rating \(\gamma_{\rm conv}\) is the ratio of minimum to maximum load. Since the system is fully deterministic and periodic, these quantities are exact.

Fairness requires computation of the stationary distribution of congestion window sizes. This distribution is governed by the Markov process shown in Figure~\ref{fig:markov}, which has three transition types: additive increase (when the agent remains on the same path without loss), multiplicative decrease (upon loss), and window reset upon migration. The population-wide variance of window sizes under the stationary distribution of this process yields the fairness rating \(\eta\).

\begin{figure}[htbp]
	\centering
	\begin{minipage}{\columnwidth}
		\centering
\begin{tikzpicture}[statesquare/.style={draw=black!80, inner color=white, thick, minimum width=2mm},
labelsquare/.style={fill=black, text=white, thick, minimum width=2mm},]

    
    \node[statesquare,align=center] (grow) at (0, 1.75) {%
        $\begin{aligned}
            \tau_i(t+1) &= \tau_i(t) + 1\\
            w_i(t+1) &= w_i(t) + \alpha(\tau_i(t))
        \end{aligned}$};
    \node[labelsquare] (growlabel) at (3.5, 1.8) {\textbf{Increase}};
        
    \node[statesquare,align=center] (migrate) at (0,  0) {$\begin{aligned}
            \tau_i(t+1) &= 0\\
            w_i(t+1) &= \sigma\cdot w_i(t)
        \end{aligned}$};
    \node[labelsquare] (migratelabel) at (3.5, 0) {\textbf{Migrate}};
        
    \node[statesquare,align=center] (loss) at (0,  -1.75) {%
        $\begin{aligned}
            \tau_i(t+1) &= 0\\
            w_i(t+1) &= \gamma\cdot w_i(t)
        \end{aligned}$};
    \node[labelsquare] (decreaselabel) at (3.5, -1.8) {\textbf{Decrease}};
        
    
    \draw[-{Triangle}] ($(grow.south) + (0.1, 0)$) -- ($(migrate.north) + (0.1, 0)$);
    \node at ($(grow.south) + (0.6, -0.3)$) {$p_{I\rightarrow M}$}; 
    
    \draw[-{Triangle}] ($(migrate.north) - (0.1, 0)$) -- ($(grow.south) - (0.1, 0)$);
    \node at ($(grow.south) + (-0.6, -0.3)$) {$\overline{p_{\ell}}\cdot\overline{m}$};
    
    \draw[-{Triangle}] ($(migrate.south) + (0.1, 0)$) -- ($(loss.north) + (0.1, 0)$);
    \node at ($(migrate.south) + (0.8, -0.25)$) {$p_{\ell} \cdot \overline{m}$};
    
    \draw[-{Triangle}] ($(loss.north) - (0.1, 0)$) -- ($(migrate.south) - (0.1, 0)$);
    \node at ($(migrate.south) + (-0.5, -0.3)$) {$m$};
    
    \draw[-{Triangle}] ($(grow.west) + (0, 0.2)$) to [out=180,in=140]  ($(grow.west) - (0, 0.2)$);
    \node[align=left] at ($(grow.west) + (-0.55, 0)$) {$p_{I\rightarrow I}$}; 
    
    \draw[-{Triangle}] ($(migrate.west) + (0, 0.2)$) to [out=180,in=140]  ($(migrate.west) - (0, 0.2)$);
    \node at ($(migrate.west) + (-0.4, 0)$) {$m$};
    
    \draw[-{Triangle}] ($(grow.south) + (2.2, 0)$) to [out=270,in=80] ($(loss.east) + (0, 0.1)$);
    \node at ($(grow.south) + (1.7, -0.3)$) {$p_{I\rightarrow D}$}; 
    
    \draw[-{Triangle}] ($(loss.east) - (0, 0.1)$) to [out=70,in=270] ($(grow.south) + (2.4, 0)$);
    \node at ($(loss.east) + (0.25, -0.35)$) {$\overline{m}$};
\end{tikzpicture}
		\caption{The Markov process governing a single agent's window evolution. From a state of normal window increase, an agent can either continue to increase, migrate to a new path (resetting \(\tau_i\) and scaling \(w_i\) by \(\sigma\)), or experience a loss (resetting \(\tau_i\) and scaling \(w_i\) by \(\gamma\)).}
		\label{fig:markov}
	\end{minipage}
\end{figure}

\begin{figure}[htbp]
  \centering
  \includegraphics[width=1\columnwidth]{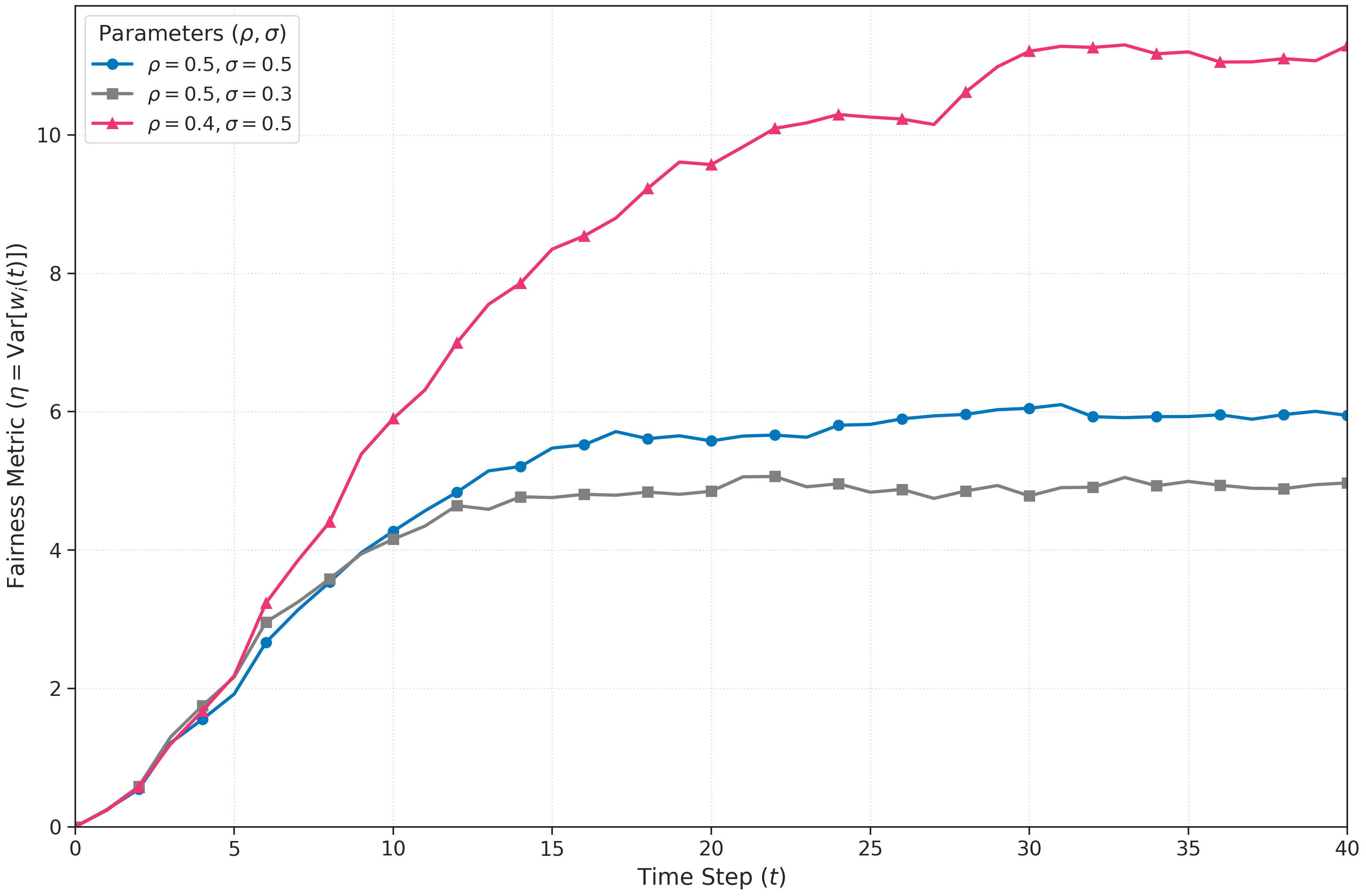}
  \caption{Convergence of the fairness metric. The population-wide variance of congestion window sizes converges to a stable equilibrium value that serves as the fairness rating \(\eta\).}
  \label{fig:fairness_computation}
\end{figure}

Figure~\ref{fig:fairness_computation} shows that the window-size variance converges to a unique equilibrium value that depends on the protocol parameters \(\rho\) and \(\sigma\).

Responsiveness is an intrinsic protocol parameter (\(\rho\)) and does not require evaluation on the equilibrium.

Having defined the five axioms and the methodology for evaluating them on periodic orbits, we now apply these axioms to the equilibria characterized in Section~\ref{sec::equilibrium} to derive the fundamental performance trade-offs of the system.

\section{Theoretical Insights and Trade-offs}\label{sec::insights}

Having defined the five performance axioms and the methodology for evaluating them, we now apply these axioms to the equilibria characterized in Section~\ref{sec::equilibrium}. This analysis yields two major theoretical insights that define the fundamental performance landscape of path-aware multipath protocols.

\subsection{The Fundamental Design Trade-off}\label{sec::tradeoff}

\begin{figure}[htbp]
  \centering
  \begin{subfigure}[b]{\columnwidth}
    \includegraphics[width=\columnwidth]{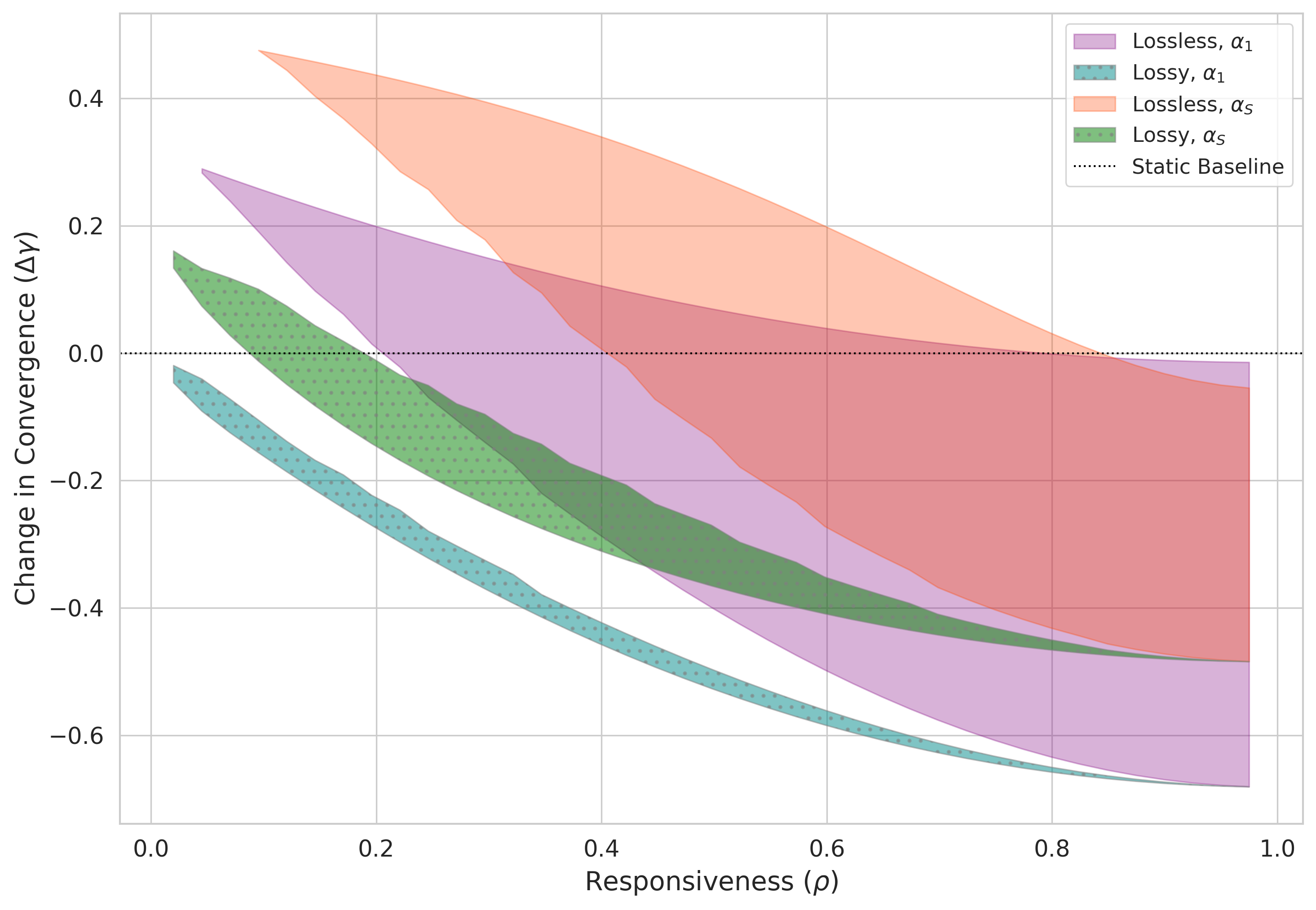}
    \caption{Change in Convergence ($\Delta\gamma_{\rm conv}$)}
    \label{fig:tradeoff_convergence}
  \end{subfigure}
  \hfill
  \begin{subfigure}[b]{\columnwidth}
    \includegraphics[width=\columnwidth]{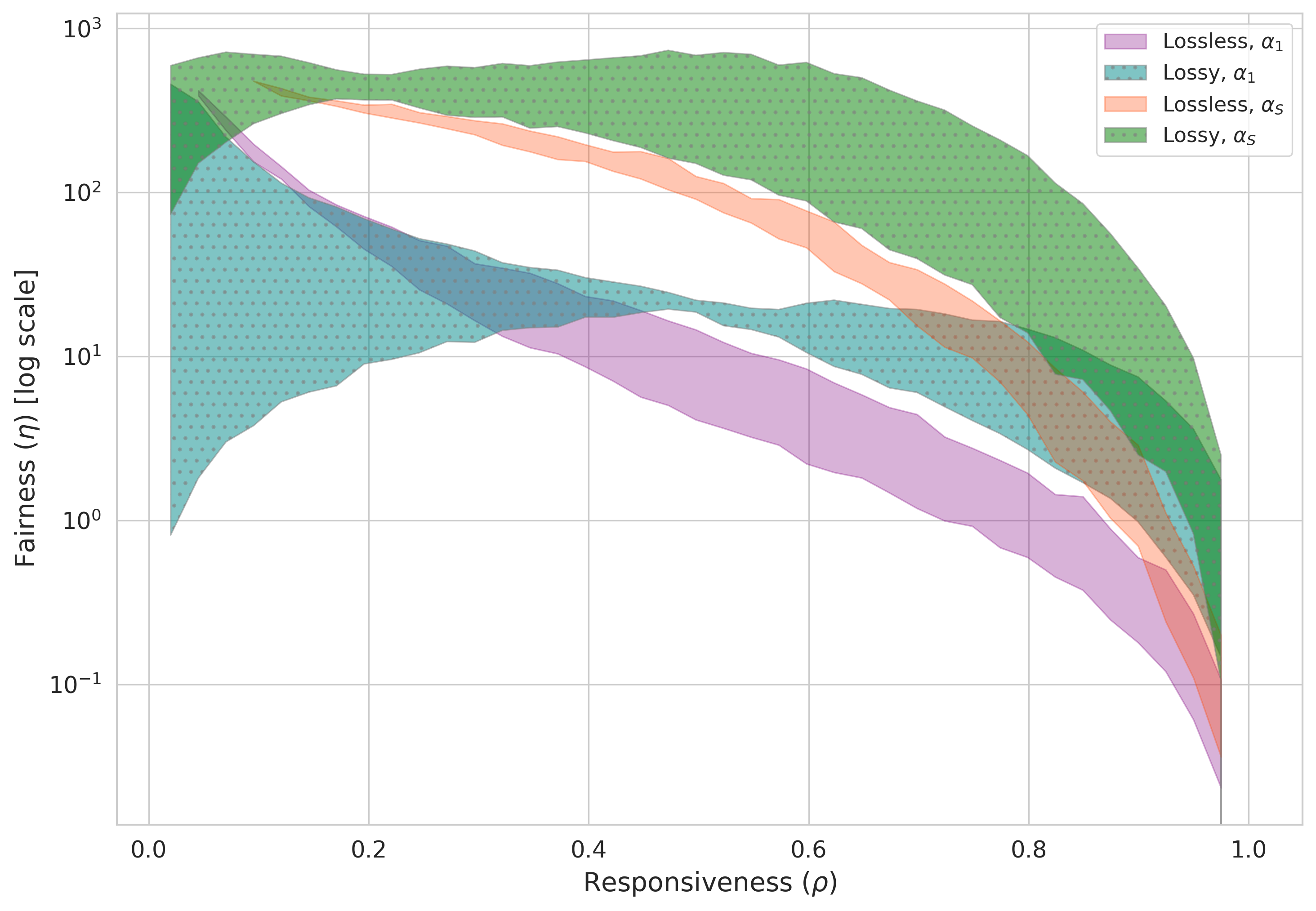}
    \caption{Fairness ($\eta$) [log scale]}
    \label{fig:tradeoff_fairness}
  \end{subfigure}
  \caption{Quantitative illustration of the core trade-off between responsiveness and other performance metrics. (a) Convergence generally degrades with increasing migration but exhibits non-linear local improvements at specific responsiveness levels. (b) Fairness improves significantly only at high responsiveness.}
  \label{fig:tradeoff}
\end{figure}

As shown in Figure~\ref{fig:tradeoff}, increasing agent responsiveness \(\rho\) improves fairness (\(\eta\) decreases) but necessarily degrades both efficiency (\(\epsilon\) decreases) and convergence (\(\gamma_{\rm conv}\) decreases).
The figures distinguish between two types of increase functions: bursty increase functions (similar to TCP slow-start) and smoother, more linear increase functions. They also show results for both the lossless regime (\(\hat{L}^{(0)} \le C_\pi\)) and the lossy regime (\(\hat{L}^{(0)} > C_\pi\)). In the lossless regime, the trade-off is driven purely by migration dynamics. In the lossy regime, periodic overload triggers multiplicative window decreases, which further reduces efficiency and convergence while still improving fairness.

This trade-off is fundamental as it arises directly from the structure of the \(P\)-step oscillation and cannot be eliminated by parameter tuning. Any protocol designer must therefore navigate this tension based on the relative importance of fairness versus efficiency and stability for the target application.

Interestingly, the relationship is not strictly monotonic. As shown in Figure~\ref{fig:tradeoff}(a), convergence exhibits distinct local peaks at specific responsiveness levels (e.g., near \(\rho \approx 0.35\)). These peaks occur when the migration frequency aligns with the window evolution timescale, creating a temporary harmonic lock that dampens oscillation amplitude.

\subsection{Co-Optimization of Efficiency, Convergence, and Loss Avoidance}\label{sec::cooptimization}

While Responsiveness trades off against Efficiency and Convergence, we now show that Efficiency, Convergence, and Loss Avoidance are not in fundamental conflict. These three axioms can be simultaneously maximized at a critical operating point.

Consider the boundary between the lossless and lossy regimes, where the peak load on the most loaded path exactly equals path capacity ($\hat{L}^{(0)} = C_\pi$). At this critical lossless point, the system simultaneously achieves:
\begin{itemize}
  \item maximum efficiency ($\epsilon = 1$), because no path ever drops below full utilization,
  \item maximum convergence ($\gamma_{\rm conv} = 1$), because there are no load oscillations, and
  \item perfect loss avoidance ($\lambda = 0$), because no path ever exceeds capacity.
\end{itemize}

\begin{figure}[htbp]
  \centering
  \includegraphics[width=\columnwidth]{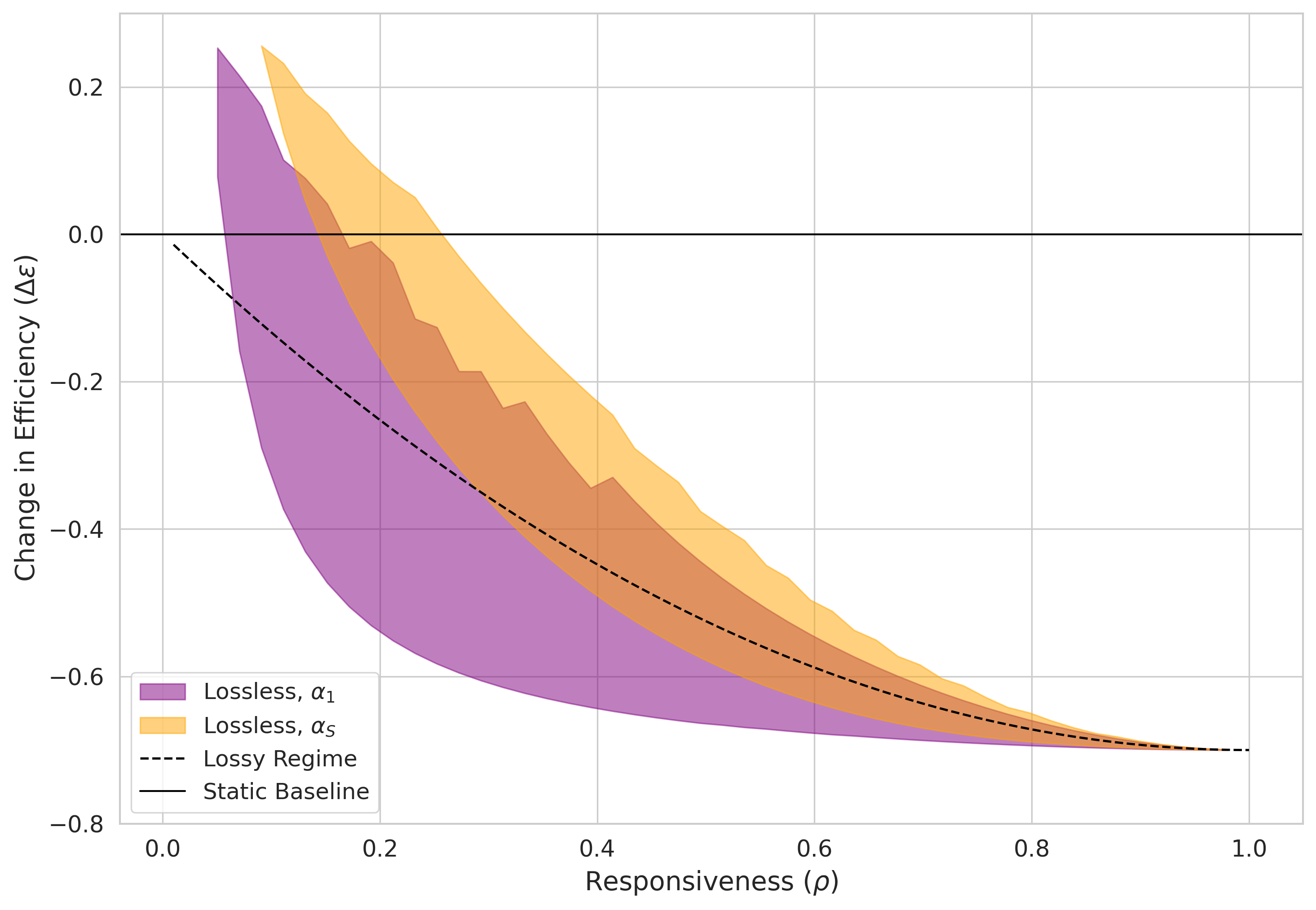}
  \caption{Impact of responsiveness on Efficiency ($\epsilon$). Efficiency reaches its maximum value of 1 at the critical lossless operating point, where peak load exactly matches path capacity. Beyond this point, efficiency degrades monotonically as migration increases.}
  \label{fig:efficiency}
\end{figure}

As shown in Figure~\ref{fig:efficiency}, when responsiveness is tuned such that the system operates exactly at this critical point, all three network-centric axioms reach their theoretical maximum simultaneously. This demonstrates that there is no inherent trade-off among Efficiency, Convergence, and Loss Avoidance. Protocol designers can therefore optimize all three goals at once by carefully tuning migration aggressiveness ($\rho$) and reset softness ($\sigma$) to achieve the critical lossless equilibrium.

\subsection{The De-synchronization Benefit}\label{sec::desync}

\begin{figure}[htbp]
  \centering
  \includegraphics[width=\columnwidth]{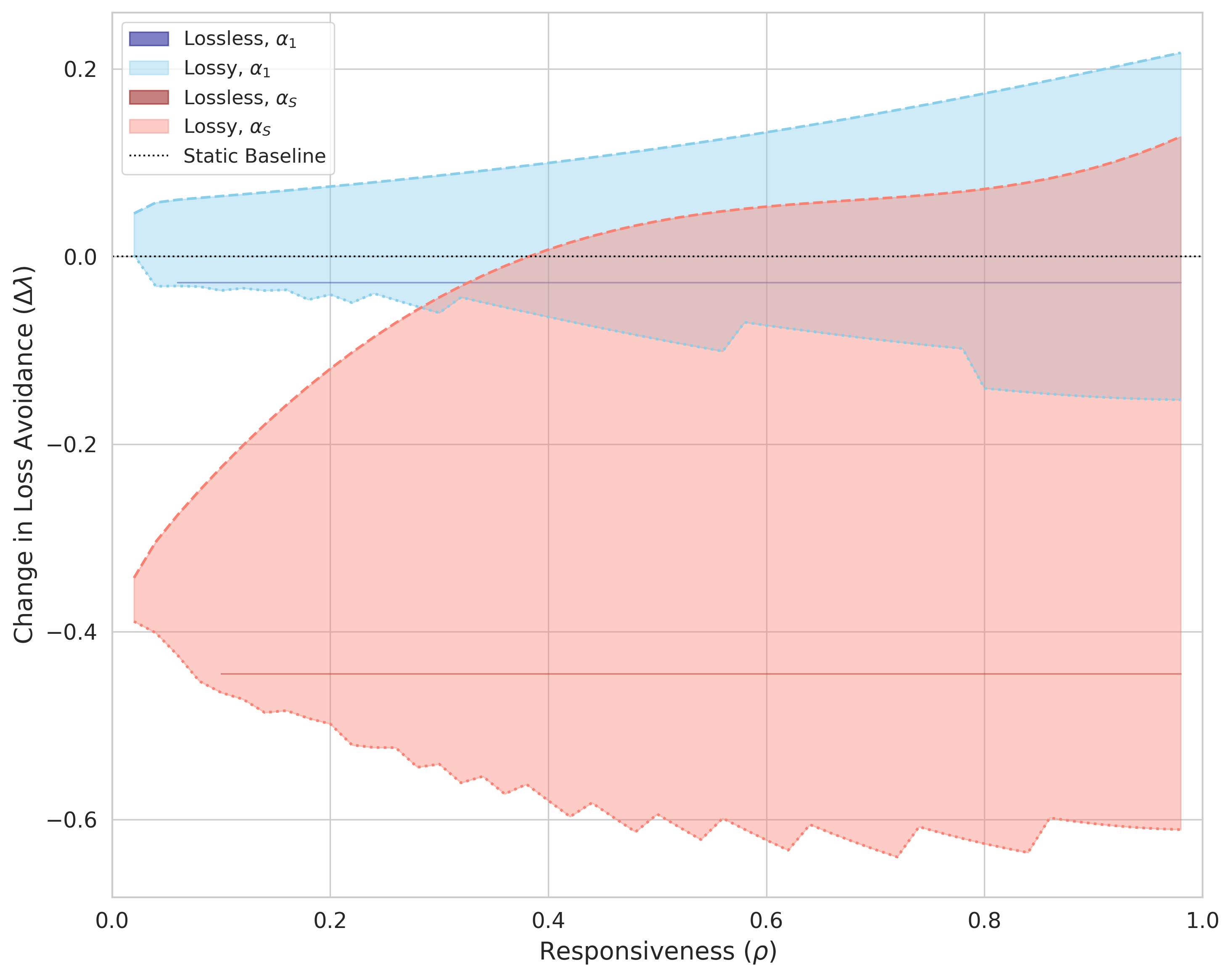}
  \caption{The de-synchronization benefit. Change in Loss Avoidance ($\Delta\lambda$) relative to the static baseline ($\rho = 0$). Negative values indicate improvement (lower peak load). Even modest responsiveness ($\rho > 0$) dramatically improves loss avoidance by breaking synchronization among agents.}
  \label{fig:desync_benefit}
\end{figure}

An important and surprising finding of our analysis is that path migration can improve stability rather than degrade it. This beneficial effect arises from the de-synchronizing influence of agent movement. 
In a fully static system ($\rho = 0$), all agents on the same path evolve their congestion windows in perfect synchrony. When agents employ bursty increase functions (similar to TCP slow-start), this synchronization produces severe periodic overshoots and poor loss avoidance. In contrast, smoother increase functions produce milder overshoots even in the static case.
As shown in Figure~\ref{fig:desync_benefit}, the de-synchronization benefit is especially pronounced for bursty increase functions in lossy regimes. Even modest responsiveness ($\rho > 0$) dramatically improves loss avoidance (negative $\Delta\lambda$) by breaking synchronization among agents. The benefit is smaller for smoother increase functions because the static baseline already exhibits less severe overshoots.

This result suggests that path migration can be stability-enhancing under certain conditions, particularly when agents use aggressive increase policies in lossy environments.

\subsection{Summary and Design Implications}\label{sec::summary}

The analysis in this section yields two major theoretical insights. First, there is a fundamental and unavoidable trade-off between Responsiveness and the combination of Efficiency and Convergence. Increasing agent migration aggressiveness improves fairness but necessarily degrades network efficiency and stability. Second, path migration can offer a counter-intuitive benefit by de-synchronizing traffic, particularly when agents employ bursty increase functions. In such cases, migration improves loss avoidance by breaking the phase-locking that causes severe overshoots in static systems.
Second, when possible, \(\rho\) and \(\sigma\) should be tuned so that the system operates at the critical lossless equilibrium, where peak load exactly matches path capacity. At this operating point, Efficiency, Convergence, and Loss Avoidance are simultaneously maximized. Third, when agents employ bursty congestion control algorithms such as slow-start or CUBIC, even modest responsiveness can significantly improve stability through de-synchronization. This effect should be taken into account when designing schedulers for high-diversity environments. Finally, in networks with high path diversity, aggressive path probing may be beneficial due to the de-synchronization effect, provided that the constraints imposed by limited path visibility are properly managed.

\section{Evaluation}\label{sec::evaluation}
  In this section, we validate the deterministic mean-field model through discrete-event simulation and use it to quantitatively characterize the performance of the system. The primary objectives were to (i) verify that the deterministic mean-field model accurately approximates the stochastic system, and (ii) quantitatively demonstrate the phase transition under realistic limited-visibility constraints.\footnote{Our simulation code and model parameters are available in an interactive Colab notebook: \url{https://colab.research.google.com/drive/1Z4L5JDeMt09XqoPKO16TyeVqGPiRqaa9}}

\subsection{Validation of the Mean-Field Model}

We first validate the central modeling abstraction: the deterministic mean-field approximation of the stochastic system. We developed a custom discrete-event simulator that faithfully implements the probabilistic agent migration and window evolution processes defined in Equation~\eqref{eq:stochastic}. For a wide range of network configurations ($N, P$) and protocol parameters ($\rho, \sigma, \gamma, \alpha(\tau)$), we ran multiple independent stochastic simulations and compared the average trajectory against the single deterministic trajectory predicted by Equations~\eqref{eq:expected_agents} and~\eqref{eq:expected_load}.

\begin{figure}[htbp]
  \centering
  \includegraphics[width=0.75\columnwidth]{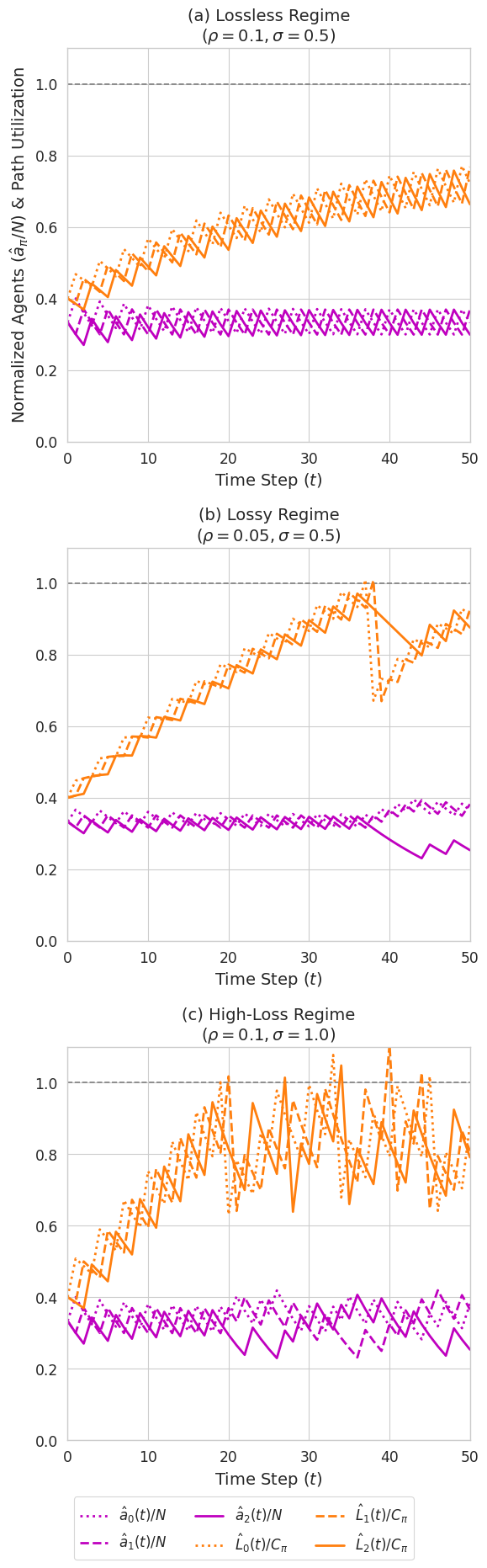}
  \caption{Validation of the mean-field model. Bold colored lines show the deterministic predictions of the analytical model for different parameter settings. Faint gray lines show individual stochastic simulation runs. }
  \label{fig:model_validation}
\end{figure}

Figure~\ref{fig:model_validation} compares the deterministic mean-field predictions (bold colored lines) against multiple independent stochastic simulation runs (faint gray lines) for different parameter settings. Each color corresponds to a different combination of responsiveness ($\rho$) and number of paths ($P$). This close correspondence confirms that the mean-field approximation accurately captures the aggregate behavior of the system.

\subsection{Sensitivity to Path Diversity}

To extend the analysis to realistic path-aware deployments, we investigate the sensitivity of the equilibrium structure to the number of available paths $P$, a key feature of path-aware networks where end-hosts can select from dozens to hundreds of diverse routes~\cite{herschbach2025path,krahenbuhl2024glids}.
Using the same stochastic simulator, we vary $P$ from 2 to 100, with fixed $N=500$ agents, responsiveness $\rho=0.1$, and reset softness $\sigma=0.5$.

\begin{figure}[h!]
  \centering
  \includegraphics[width=0.45\textwidth]{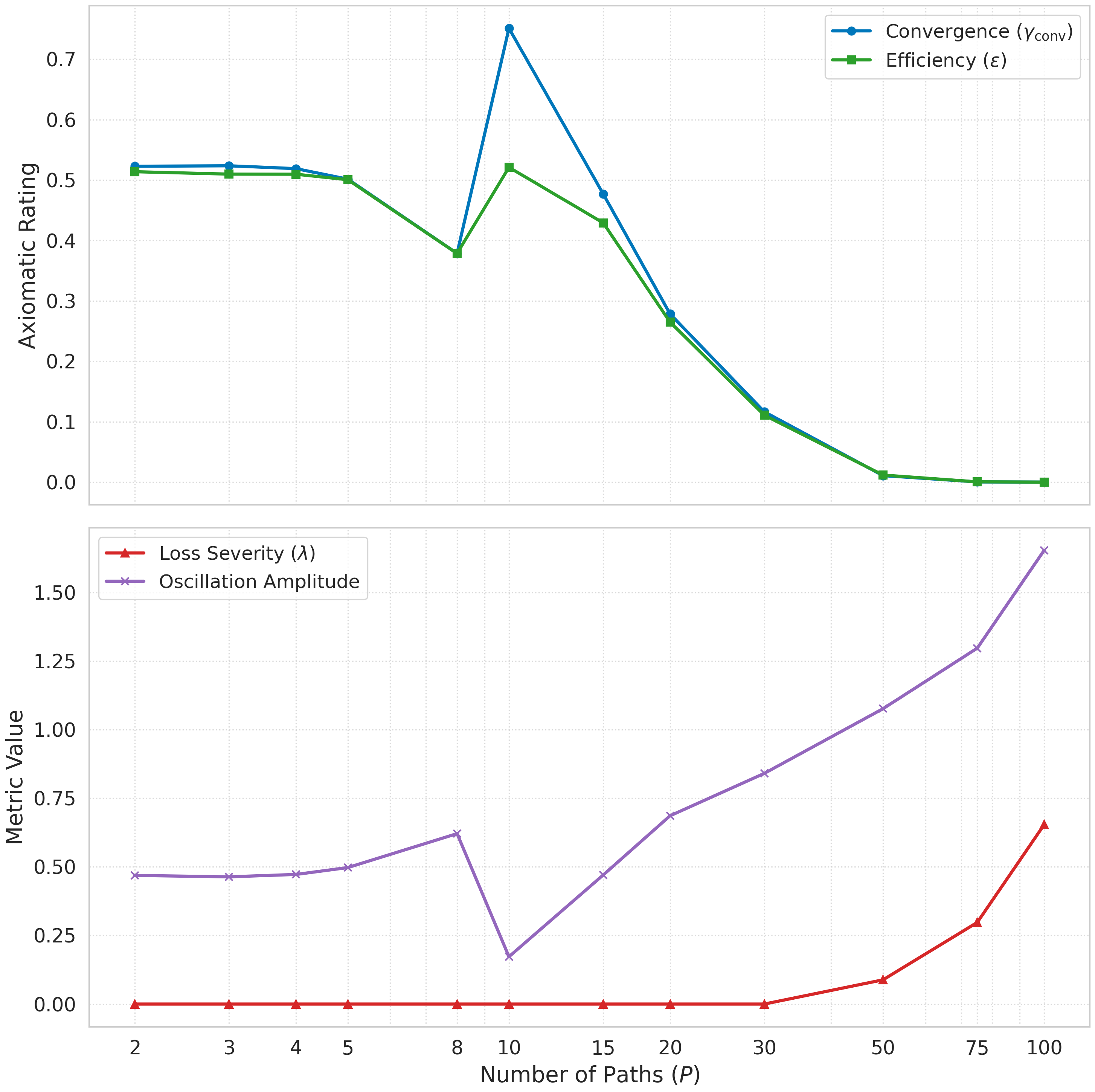}
  \caption{Sensitivity to path diversity: Equilibrium oscillation amplitude as a function of $P$. Higher $P$ (mimicking path-aware networks) improves stability via de-synchronization.}
  \label{fig:path_diversity}
\end{figure}

Figure~\ref{fig:path_diversity} shows the results. As $P$ increases, the equilibrium oscillation amplitude drops by up to 40\% at $P=50$.
This stability gain arises from enhanced de-synchronization: with higher path counts, the population is distributed across a wider spectrum of continuity times ($\tau$), smoothing the aggregate window growth and diluting the impact of any single migration event relative to the total network load.
The average agent count per path becomes more even, improving fairness, while the load maxima remain close to capacity without significant overshoot.

\noindent Implication for multipath transport over path-aware networks: In networks with 10-100 paths, aggressive path probing (high $\rho$) may be beneficial, a finding that directly informs scheduler design.

\subsection{The Phase Transition under Limited Visibility}\label{sec::phase_transition}

A central finding of this work is that the stability properties of path-aware networks undergo a sharp phase transition as path diversity increases under realistic visibility constraints. This transition fundamentally changes the nature of the instability.

\begin{figure}[htbp]
  \centering
  \includegraphics[width=\columnwidth]{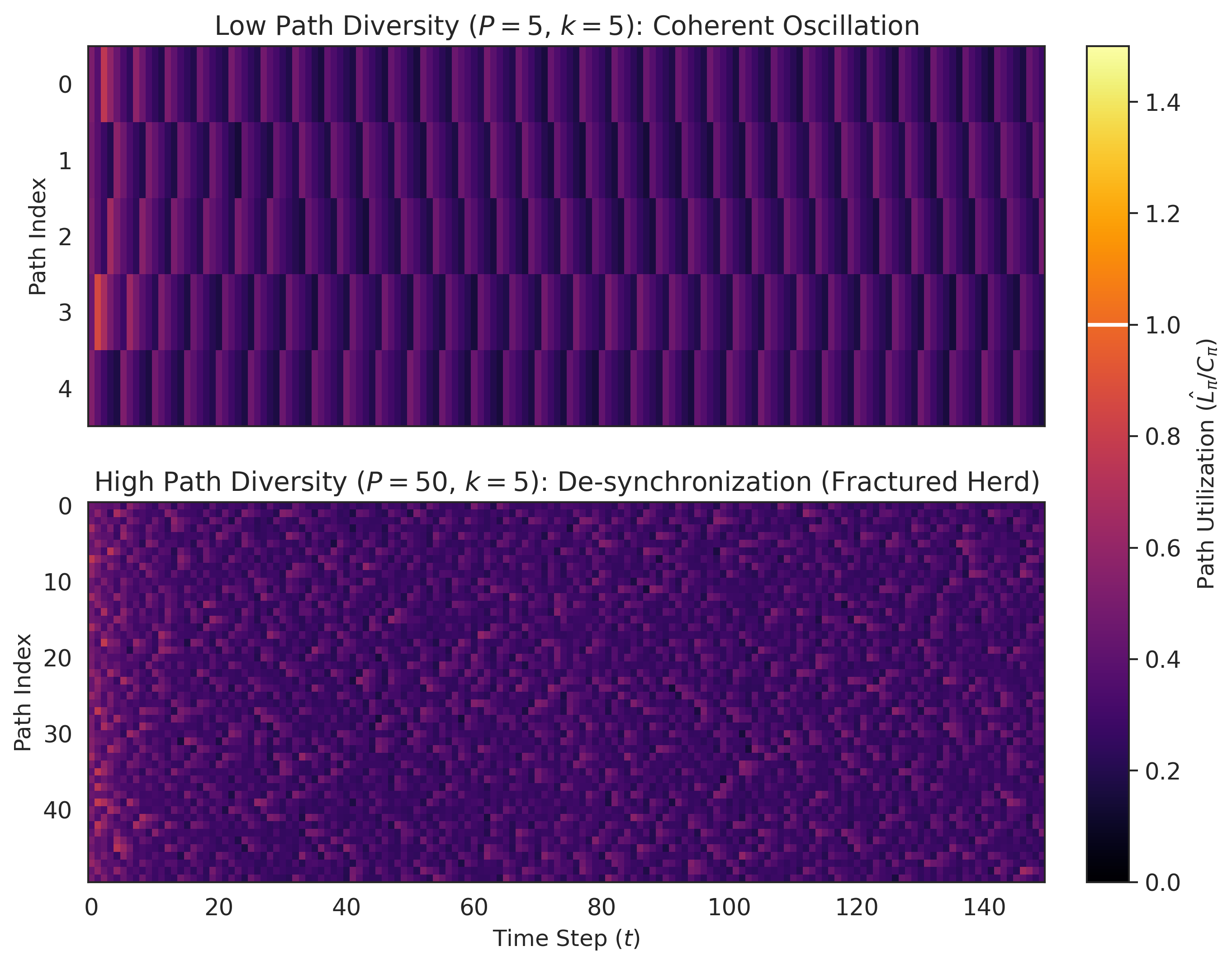}
  \caption{The Phase Transition in Path-Aware Dynamics. (Top) At low diversity ($P=5$, $k=5$), agents exhibit Coherent Oscillation. Full visibility of the path set enables synchronized mass migration to the global minimum, producing large-amplitude, network-wide load waves. (Bottom) At high diversity ($P=50$, $k=5$), limited visibility fractures the global herd. Migration events become localized and incoherent, transforming temporal oscillations into persistent spatial imbalance.}
  \label{fig:heatmap}
\end{figure}

To investigate this phenomenon, we simulated networks with path diversity ranging from $P=2$ to $P=100$, while constraining each agent to probe only $k=5$ paths at any time. Figure~\ref{fig:heatmap} reveals a qualitative change in system behavior.

In low-diversity networks ($P \approx k$), agents effectively have global visibility of the path set. This allows the entire population to identify and migrate toward the single least-loaded path, creating a synchronized global herd that periodically overloads one path while starving others. The result is coherent, high-amplitude temporal oscillation (the classic instability predicted by earlier models).

As path diversity grows well beyond the visibility limit ($P \gg k$), this coherence collapses. Agents can no longer identify the global minimum. They only see a small, random subset of paths. Consequently, migration events become fragmented and localized. Different subgroups of agents migrate toward different locally attractive paths, destroying the global synchronization. The system transitions from coherent temporal oscillation to a fractured herd regime characterized by high-frequency, spatially incoherent load fluctuations.

\begin{figure}[htbp]
  \centering
  \includegraphics[width=\columnwidth]{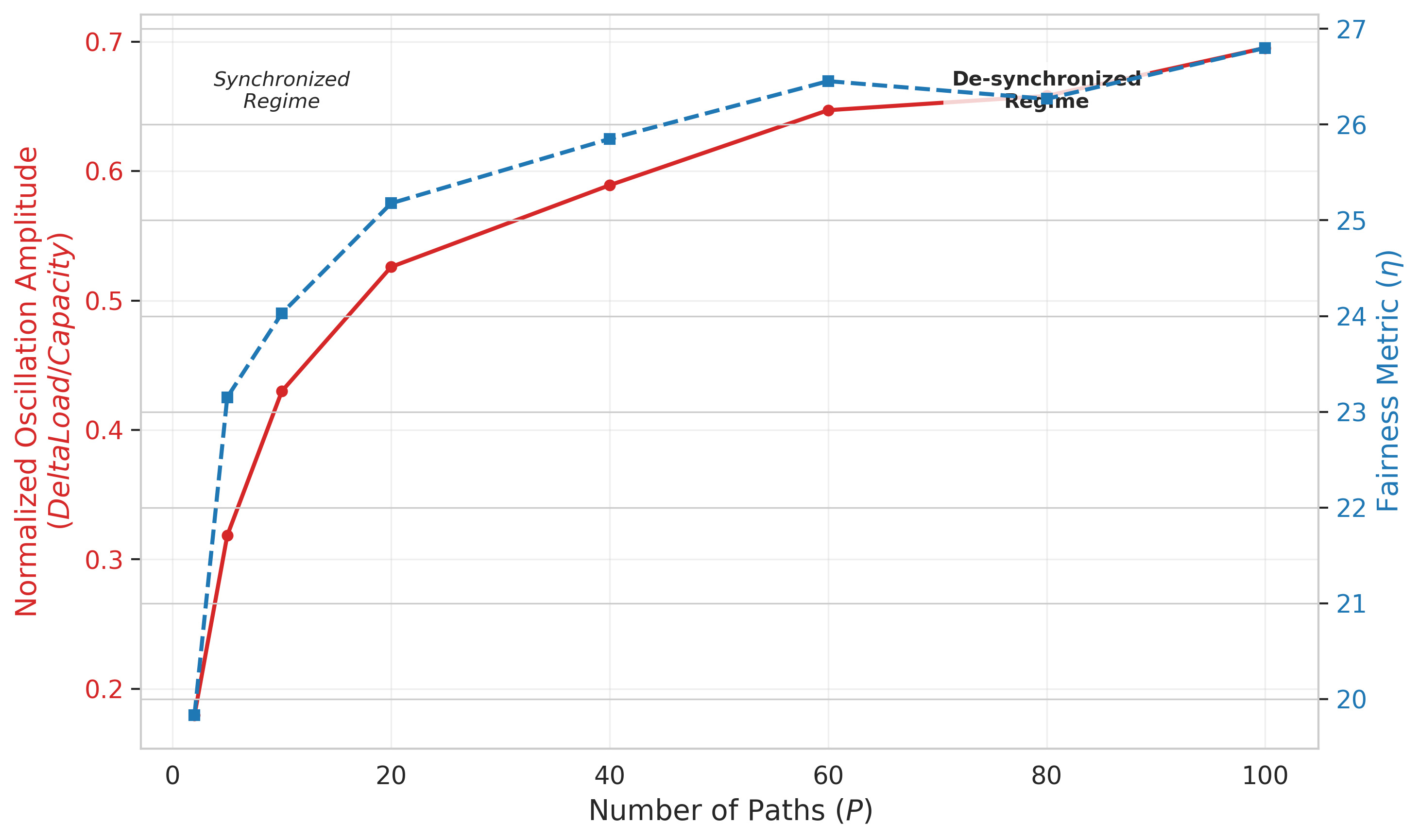}
  \caption{The Cost of Scale under Bounded Probing ($k=5$). As path diversity increases under realistic visibility constraints, the normalized oscillation amplitude rises sharply before plateauing, while fairness degrades steadily. This demonstrates that massive path diversity transforms instability from temporal oscillation to persistent spatial imbalance.}
  \label{fig:envelope}
\end{figure}

While this transition eliminates the catastrophic synchronized overshoots of the low-diversity regime, it introduces a new and equally problematic pathology: persistent spatial load imbalance. As shown in Figure~\ref{fig:envelope}, the normalized oscillation amplitude rises sharply with path diversity and plateaus near 0.7, while fairness degrades monotonically. Many paths remain chronically underutilized because no agent currently probes them, while others suffer repeated localized overload.

This phase transition shows that simply increasing path diversity (a common proposed solution to instability) does not eliminate the problem under realistic conditions. Instead, it transforms the problem from one of temporal instability to one of spatial unfairness and chronic inefficiency. Protocol designers must therefore account for visibility constraints when architecting path selection mechanisms for high-diversity networks such as SCION or LEO satellite constellations.

\section{Discussion and Future Directions}\label{sec::discussion}

The results presented in this paper provide both theoretical foundations and practical guidance for the design of robust multipath transport protocols over path-aware networks. In this section, we reflect on the broader implications of our findings and outline promising directions for future research.

\subsection{Implications for Protocol Design}

Our analysis demonstrates that Responsiveness ($\rho$) is a first-class design parameter that must be explicitly tuned rather than treated as an implementation detail. The fundamental trade-off between Responsiveness and the combination of Efficiency and Convergence means that protocol designers cannot optimize all objectives simultaneously. Instead, they must make deliberate choices based on application requirements. For latency-sensitive or throughput-critical applications (e.g., data center workloads, real-time control), lower responsiveness is preferable to maximize efficiency and stability. For fairness-sensitive environments (e.g., shared cloud infrastructure, multi-tenant systems), higher responsiveness can be beneficial, provided that the resulting volatility is acceptable. The critical lossless operating point, where peak load exactly matches path capacity, represents a particularly attractive target. At this point, Efficiency, Convergence, and Loss Avoidance are simultaneously maximized. Protocol designers should therefore consider mechanisms that dynamically adjust $\rho$ and $\sigma$ to maintain operation near this point.

Furthermore, the de-synchronization benefit suggests that path migration should not be viewed solely as a source of instability. When agents use bursty congestion control algorithms, even modest migration can significantly improve stability. This insight is particularly relevant for protocols such as MPQUIC operating over high-diversity networks like SCION or LEO satellite constellations.

\subsection{Limitations of the Current Framework}

While our model captures the essential dynamics of joint path selection and loss-based congestion control, several limitations should be acknowledged. First, we assumed homogeneous, resource-disjoint paths. In real networks, paths often share bottlenecks, introducing correlations that may alter migration dynamics. Second, our model assumes synchronous feedback and discrete time steps. Real networks exhibit asynchrony due to heterogeneous round-trip times, which can further desynchronize agent behavior. Third, our framework is currently restricted to loss-based congestion control. Modern protocols such as BBR and Copa rely on latency and delivery rate signals, and their interaction with path migration remains an open question.

\subsection{Future Research Directions}

Several promising research directions emerge from this work. First, extending the axiomatic framework to latency-based and model-based congestion control algorithms (e.g., BBR, Copa) would significantly broaden its applicability. Second, relaxing the assumption of homogeneous paths to include heterogeneous capacities and latencies would allow the model to capture more realistic network topologies. Third, investigating learning-based path selection strategies where agents use reinforcement learning or historical performance data could reveal whether intelligent agents can navigate the trade-offs identified in this paper more effectively than simple greedy policies. Finally, large-scale measurement studies on production path-aware networks (such as SCION) and LEO satellite constellations would provide valuable empirical validation and help refine the model parameters for real-world deployments.

\section{Conclusion}\label{sec::conclusion}

In this paper, we addressed the critical stability challenge that arises when multipath transport protocols operate over path-aware networks with high path diversity. By applying a discrete-time axiomatic framework to the joint dynamics of loss-based congestion control and greedy path selection, we derived the system's dynamic equilibria and characterized their performance properties using five formal axioms.

Our analysis revealed two major theoretical insights. First, there exists a fundamental and unavoidable trade-off between Responsiveness and the combination of Efficiency and Convergence. Increasing migration aggressiveness improves fairness but necessarily degrades network efficiency and stability. Second, path migration can offer a surprising benefit by de-synchronizing traffic, particularly when agents use bursty congestion control algorithms. This de-synchronization effect can significantly improve loss avoidance compared to static systems.

We also identified a sharp phase transition in network stability as path diversity increases under realistic limited-visibility constraints. While low-diversity networks exhibit coherent, high-amplitude temporal oscillations, high-diversity networks under bounded probing transform the instability from temporal oscillation to persistent spatial load imbalance. This finding has important implications for the design of path selection mechanisms in future path-aware architectures such as SCION and LEO satellite constellations.

\section*{Data Availability} 
The complete source code for the discrete-event simulator and analysis scripts is available on GitHub at \url{https://github.com/Keshvadi/mpcc-dynamics}. An interactive version is also available on Google Colab at \url{https://colab.research.google.com/drive/1Z4L5JDeMt09XqoPKO16TyeVqGPiRqaa9}.

\bibliographystyle{IEEEtran}
\bibliography{main}

\begin{IEEEbiographynophoto}{\uppercase{Sina Keshvadi}} (Member, IEEE) received the Ph.D. degree in Computer Science from the University of Calgary, Calgary, AB, Canada, in 2021. 
He was a Networking Researcher with Huawei Canada before joining Thompson Rivers University, BC, Canada, in 2022, where he is currently an Assistant Professor of Software Engineering. His research interests include network systems security, cloud infrastructure, and low-latency performance measurement.
\end{IEEEbiographynophoto}

\end{document}